\documentclass[11pt,a4paper]{article}
\pdfoutput=1
\usepackage{jcappub}

\usepackage{amsmath,amssymb,bm,mathtools}
\usepackage{graphicx}
\graphicspath{{final_figure_pack/}}
\usepackage{subfig}
\usepackage[T1]{fontenc}
\usepackage{lmodern}
\usepackage{microtype}
\usepackage{booktabs}
\usepackage{array}
\usepackage{float}
\allowdisplaybreaks

\newcommand{\vev}[1]{\overline{#1}}
\newcommand{\dd}{\mathrm{d}}
\newcommand{\ii}{\mathrm{i}}

\newcommand{\Order}[1]{\mathcal{O}\!\left(#1\right)}
\newcommand{\posteriorcode}{\texttt{analysis/}\allowbreak\texttt{cosmology\_mock\_posterior.py}}
\newcommand{\ringdowncode}{\texttt{analysis/}\allowbreak\texttt{ringdown\_bayesian\_injections.py}}

\title{A cosmology-to-ringdown EFT consistency map for scalar-tensor gravity}

\author[a]{Mushtaq Ahmad,}
\author[b]{and M. Farasat Shamir}

\affiliation[a]{National University of Computer and Emerging Sciences, Chiniot-Faisalabad Campus, Pakistan}
\affiliation[b]{School of Computing and Mathematical Sciences, University of Leicester, United Kingdom}

\emailAdd{mushtaq.sial@nu.edu.pk}
\emailAdd{farasat.shamir@leicester.ac.uk}

\abstract{We construct an effective-field-theory bridge from late-time scalar-tensor cosmology to black-hole ringdown observables. Starting from a cosmology-conditioned EFT posterior, we lift Jordan-frame FLRW data through a finite covariant jet, transport the result to the arbitrary-background EFT for black-hole perturbations with a timelike scalar, and project it onto parity-resolved quasinormal-mode response kernels. The cosmological layer is a deterministic compressed likelihood built from BAO-like distances, growth summaries, low-redshift tensor-speed information, stability filters, and posterior samples for the ringdown pushforward. The detector layer uses Bayesian time-domain injections, one-, two-, and three-mode recovery models, analytic marginalization over linear sine/cosine amplitudes, remnant-calibration covariance products, and start-time variations. The transported posterior shows that FLRW tensor-speed deformations inherited from cosmology are driven far below ringdown detectability, whereas operators that vanish on homogeneous FLRW backgrounds can remain active in the anisotropic near zone of a black hole. For a literature-calibrated Hayward branch, we specify the prior measure, separate directly admissible points from a proxy continuation, and propagate both to detector-whitened consistency modes. The resulting framework turns cosmological viability into black-hole spectroscopy priors while keeping the strong-field completion explicit rather than assumed.}


\notoc

\begin{document}
\maketitle

\section{Introduction}
\label{sec:intro}

Understanding why the Universe is accelerating remains one of the clearest motivations for testing gravity beyond general relativity on the largest observable scales. Within that broader effort, single-field scalar-tensor theories are especially useful because they capture a wide range of late-time modified-gravity scenarios within a common effective description that can be confronted directly with cosmological data \cite{CliftonEtAl2012,Gubitosi2013,Bloomfield2013,BelliniSawicki2014,FrusciantePerenon2020}. Over the last decade the effective field theory (EFT) of dark energy has matured from a formal parametrization into a practical interface between covariant theory space and data analysis. Current combinations of CMB, BAO, supernova, clustering, and weak-lensing measurements already compress the viable EFT/Horndeski space strongly toward the general-relativistic limit, even though small departures can remain statistically admissible in some extended background models \cite{ZhengEtAl2025,StolznerEtAl2026}.

Gravitational-wave observations of compact-binary mergers now probe the same underlying question in a very different regime. The post-merger ringdown is controlled by the quasinormal-mode (QNM) spectrum of the remnant and underlies the program of black-hole spectroscopy and no-hair tests \cite{Dreyer2004,BertiCardosoWill2006}. The latest LIGO-Virgo-KAGRA remnant analyses, including the GWTC-4.0 remnants study and the dedicated spectroscopy analysis of GW250114, remain consistent with Kerr while also showing how quickly the constraining power of high-SNR ringdown data is improving \cite{AbbottEtAl2025,LVKGWTC4Remnants2026,LVKGW2501142025}. The outlook becomes even stronger with improved post-merger waveform modeling and with next-generation detectors such as Einstein Telescope and LISA \cite{BertiEtAl2025Review,MaggioreEtAl2020,AmaroSeoaneEtAl2017,PitteEtAl2024,BhagwatEtAl2022}. At the same time, recent reviews make clear that precision spectroscopy must contend with real theoretical and data-analysis issues --- including mode resolvability, nonlinear mode content, and spectral stability --- before it can become a precision test of gravity \cite{DestounisDuque2024,BertiEtAl2025Review}.

On the theory side, a large literature has computed QNMs in specific beyond-GR models, and recent work has begun to organize those calculations through perturbative spectral expansions and EFT descriptions of rotating black-hole ringdown \cite{ChungYunes2024,MaenautEtAl2024}. These developments are important, but they mostly answer a forward question: once a strong-field theory is chosen, what ringdown signal follows? If cosmological constraints are taken seriously, the more relevant inverse question is different: if a modified-gravity model survives late-time cosmology, what black-hole ringdown signatures remain allowed? Recent solution-specific studies sharpen this point further. One finds that dynamical dark-energy hair can generate order-unity QNM shifts in a stable scalar-tensor example, while related work has clarified the existence of regular and stable cosmological-hair branches. Together, these results make the cosmology--ringdown interface timely and show why individual solutions should be complemented by a theory-agnostic EFT consistency test \cite{SmuldersNollerSirera2026,SmuldersNoller2026}.

This question has become tractable because the black-hole side of the EFT infrastructure is now in place. An arbitrary-background EFT of black-hole perturbations with a timelike scalar profile has been formulated for scalar-tensor theories; its odd-parity sector has been reduced to a generalized Regge--Wheeler equation; the first QNM applications have been carried out; and a Jordan-frame to almost-Einstein-frame bridge has been derived that explicitly connects dark-energy EFT coefficients to black-hole EFT coefficients. Recent work has also begun to open the even-parity sector \cite{MukohyamaYingcharoenrat2022,MukohyamaTakahashiYingcharoenrat2022,MukohyamaTakahashiTomikawaYingcharoenrat2023,MukohyamaSerailleTakahashiYingcharoenrat2025,MukohyamaTakahashiTomikawaYingcharoenrat2025}. Parallel developments in vector-tensor gravity reinforce the same program, although the present analysis focuses on the timelike-scalar sector \cite{AokiEtAl2023,TomizukaEtAl2025}.

Cosmology and ringdown therefore constrain different projections of the same candidate theory. FLRW data see homogeneous and isotropic combinations, while black-hole perturbations can also see anisotropic structures that vanish identically on FLRW\@. Post-GW170817 luminality on cosmological backgrounds is a powerful constraint, but it does not by itself guarantee GR-like propagation in the near zone of a black hole \cite{EzquiagaZumalacarregui2017,CreminelliVernizzi2017,MukohyamaSerailleTakahashiYingcharoenrat2025}. A meaningful comparison must carry the cosmology-conditioned EFT measure through a covariant lift, into the black-hole EFT, and finally into the variables measured by spectroscopy. The most complete quantitative control is presently available in the odd-parity, nonspinning sector, where the arbitrary-background EFT, the master-equation reduction, and published timelike-scalar QNM calculations are mature. Two controlled extensions are also included: a weak-mixing even-sector factorization and a slow-spin continuation of the inherited branch.

This work connects these pieces at the level of posterior support. A filtered cosmological viability measure is lifted through a finite set of covariant jets, split into FLRW-visible and FLRW-silent directions, projected onto black-hole EFT operators, and tested with detector-weighted consistency modes. The lift is validated on a nontrivial $N=1$ toy jet with an explicit FLRW null direction; qualitative detector bars are replaced by Bayesian multi-mode injections and start-time systematics; one odd-parity branch is carried analytically from cosmological luminality to $\delta\omega_{220}$; and the anisotropy-activated odd sector is calibrated against published timelike-scalar QNM calculations. Ringdown deviations are then assessed against the part of spectroscopy space that remains compatible with cosmology and with the declared parent EFT class.

\begin{figure}[t]
\centering
\subfloat[From cosmological viability to ringdown observables.\label{fig:consistency_map}]{%
  \includegraphics[width=0.485\textwidth]{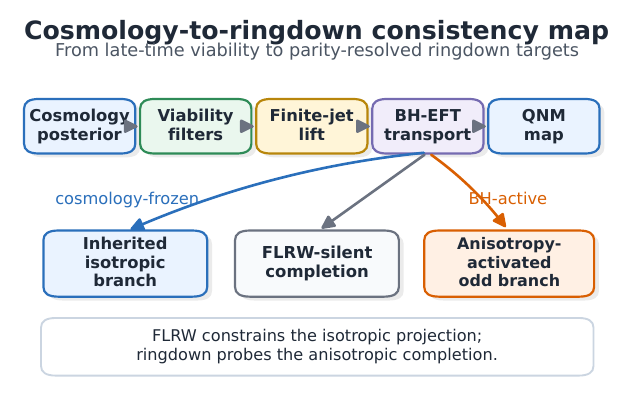}}
\hfill
\subfloat[Why cosmological luminality does not force GR-like black-hole propagation.\label{fig:luminality_loophole}]{%
  \includegraphics[width=0.485\textwidth]{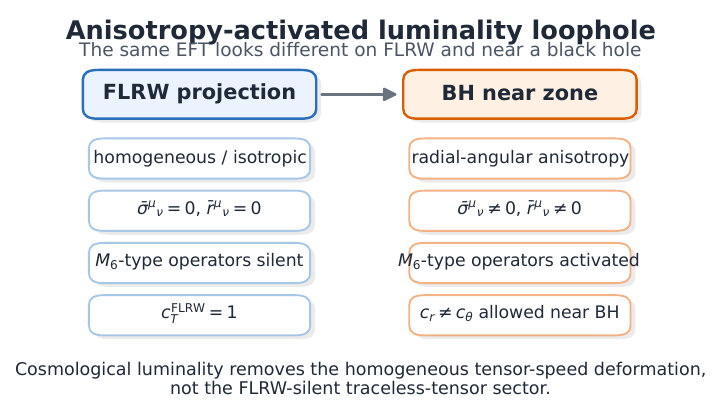}}
\caption{Core logic of the analysis. Panel (a) tracks the flow from cosmological data and viability filters to a finite-jet parent description, then to parity-resolved black-hole perturbations and allowed regions in spectroscopy space. Panel (b) shows the anisotropy-activated luminality loophole: traceless background tensors vanish on FLRW, but the same structures can be nonzero near a static black hole and modify odd-parity propagation without violating the homogeneous luminality prior.}
\label{fig:overview_loophole}
\end{figure}

Figure~\ref{fig:overview_loophole} summarizes the logic. Panel~(a) follows the flow from a filtered cosmological posterior to parity-resolved ringdown observables, while panel~(b) isolates the loophole: operators multiplying traceless background tensors are silent on FLRW but need not vanish near a black hole. The following sections develop this split between the isotropic cosmological projection and the anisotropic near-zone completion.

\subsection{Guide to quantitative claims}
\label{subsec:provenance}

Tables~\ref{tab:provenance} and~\ref{tab:provenance_ext} keep the quantitative claims on a single footing. They distinguish exact end-to-end results within a specified subclass, implemented numerical demonstrations, structural EFT statements, and literature-calibrated proxies used to anchor the anisotropy-activated sector.

The inherited stealth-Schwarzschild branch in Sec.~\ref{sec:benchmarks} gives the cleanest analytic chain from cosmological prior to \(\delta\omega_{220}\). The finite-jet lift is tested in Sec.~\ref{subsec:lift_demo} on an underdetermined \(N=1\) example with an FLRW null stripe and a tracked near-horizon completion coefficient. The detector layer is supplied by the Bayesian multi-mode injection/recovery suite, the detector-covariance products, and the start-time envelope summarized in Table~\ref{tab:ringdown_bayes} and Figs.~\ref{fig:bayes_ringdown_posteriors}--\ref{fig:start_time_systematics}. Proposition~1 is structural; the Hayward results in Sec.~\ref{subsec:lit_calibrated_odd} provide the size and direction of one calibrated odd-sector response. Within that branch, only the \(\hat\sigma=0.2\) point lies inside the demonstrated positive stable asymptotically luminal interval, while the larger positive points are used only as a proxy continuation.

\begin{table}[H]
\caption{Provenance and scope of the core quantitative claims.}
\label{tab:provenance}
\centering
\footnotesize
\begin{tabular}{p{0.24\textwidth} p{0.31\textwidth} p{0.35\textwidth}}
\hline
Result & Provenance & Scope and non-claim \\
\hline
Eqs.~\eqref{eq:qnm_scaling_stealth}--\eqref{eq:inherited_bound_final} and Table~\ref{tab:stealth220} & Exact odd-parity stealth-Schwarzschild rescaling plus direct-integration cross-check from Ref.~\cite{MukohyamaTakahashiTomikawaYingcharoenrat2023}. & End-to-end inherited branch; not claimed outside the adopted odd-sector subclass. \\
Table~\ref{tab:posterior_summary}, Fig.~\ref{fig:cosmo_posterior_pushforward}, and \texttt{data/} & Executable compressed mock likelihood with distance, growth and tensor-speed summaries, stability filters and posterior samples. & Posterior object used for the transport, based on simulated compressed data; a full EFTCAMB/hi\_class chain can replace it. \\
Eqs.~\eqref{eq:toy_lift_matrix}--\eqref{eq:toy_M6_predictor} and Fig.~\ref{fig:lift_demo} & Explicit $N=1$ lift with an FLRW null stripe and a near-horizon predictor. & Regularized underdetermined lift; Sec.~\ref{subsec:implemented_cosmo_posterior} supplies the sampled posterior used for the pushforward. \\
Proposition~1 and Eqs.~\eqref{eq:alphat_hayward}--\eqref{eq:nearzone_split} & Structural arbitrary-background EFT result plus the asymptotically luminal example of Ref.~\cite{MukohyamaSerailleTakahashiYingcharoenrat2025}. & Establishes the anisotropy-activated loophole; does not assign its posterior weight in every parent theory class. \\
\hline
\end{tabular}
\end{table}

\begin{table}[H]
\caption{Provenance and scope of the literature-calibrated, detector-side and extension claims.}
\label{tab:provenance_ext}
\centering
\footnotesize
\begin{tabular}{p{0.24\textwidth} p{0.31\textwidth} p{0.35\textwidth}}
\hline
Result & Provenance & Scope and non-claim \\
\hline
Eq.~\eqref{eq:response_eta}, Table~\ref{tab:hayward220}, and Fig.~\ref{fig:hayward_plane} & Literature-calibrated Hayward odd-parity QNM response from Ref.~\cite{MukohyamaTakahashiTomikawaYingcharoenrat2023}. & The $\hat\sigma=0.2$ point is inside the demonstrated stable interval; larger points are proxy continuation. \\
Table~\ref{tab:ringdown_bayes}, Eq.~\eqref{eq:prior_models_eta}, and Figs.~\ref{fig:bayes_ringdown_posteriors}--\ref{fig:detectability_hierarchy} & Bayesian time-domain injection suite with one-, two-, and three-mode recovery, analytic amplitude marginalization, remnant-calibration covariance products, and start-time variation. & Detector covariance and systematic check for the benchmark; not a full population-level LVK/LISA inference. \\
Eqs.~\eqref{eq:slowspin_factorization}--\eqref{eq:slowspin_inherited_bound} & Controlled slow-spin continuation of the inherited branch, motivated by Kerr-EFT progress \cite{MaenautEtAl2024,BoyceSantos2026}. & Extends the frozen inherited null result; does not solve axisymmetric anisotropy-activated transport. \\
Proposition~2, Eqs.~\eqref{eq:even_schur}--\eqref{eq:even_weakmix_shift}, and App.~\ref{app:extensions} & Weak-mixing even-sector derivation based on Ref.~\cite{MukohyamaTakahashiTomikawaYingcharoenrat2025}. & Protects gravitational-led even modes from linear scalar-metric contamination; not a full even-sector posterior pushforward. \\
\hline
\end{tabular}
\end{table}

\subsection{Relation to prior work and novelty}
\label{subsec:novelty}

Most ingredients used here have clear precedents. The EFT of dark energy supplies the cosmological language, arbitrary-background black-hole EFT supplies the strong-field perturbation language, published timelike-scalar QNM calculations calibrate a concrete odd-sector response, and standard spectroscopy likelihood methods provide the detector metric \cite{Gubitosi2013,Bloomfield2013,BelliniSawicki2014,FrusciantePerenon2020,MukohyamaYingcharoenrat2022,MukohyamaTakahashiYingcharoenrat2022,MukohyamaTakahashiTomikawaYingcharoenrat2023,BertiCardosoWill2006,MaggioreEtAl2020,AmaroSeoaneEtAl2017}. The new element is the combined posterior transport: a cosmology-conditioned posterior is lifted through a finite covariant jet, decomposed into FLRW-image and FLRW-silent directions, transported to black-hole EFT coefficients, and whitened by a detector covariance in QNM space. Table~\ref{tab:novelty_matrix} summarizes that separation.

\begin{table}[H]
\caption{Novelty matrix. Previous work supplies the main component theories and methods; the new contribution is their posterior-level transport, together with the explicit separation between inherited FLRW tensor-speed shifts and FLRW-silent anisotropy-activated near-zone operators.}
\label{tab:novelty_matrix}
\centering
\small
\begin{tabular}{p{0.23\textwidth} p{0.31\textwidth} p{0.34\textwidth}}
\hline\hline
Ingredient & Established in prior work & Added in this paper \\
\hline
Cosmological scalar-tensor EFT & EFT-of-dark-energy and Bellini-Sawicki property-function descriptions of the homogeneous and linear-perturbation sector \cite{Gubitosi2013,Bloomfield2013,BelliniSawicki2014,FrusciantePerenon2020}. & A reproducible compressed posterior layer, including mock data, stability filters, posterior samples, and a deterministic pushforward to ringdown variables. \\
Black-hole perturbation EFT & Arbitrary-background EFT with a timelike scalar profile and odd/even perturbation reductions \cite{MukohyamaYingcharoenrat2022,MukohyamaTakahashiYingcharoenrat2022,MukohyamaTakahashiTomikawaYingcharoenrat2025}. & A finite-jet transport interface that identifies which covariant directions are fixed by FLRW data and which remain FLRW-silent but black-hole active. \\
Frame bridge and luminality & The Jordan/almost-Einstein frame bridge and explicit asymptotically luminal black-hole EFT examples \cite{MukohyamaSerailleTakahashiYingcharoenrat2025}. & A no-go style inherited-branch result showing that cosmological luminality freezes homogeneous tensor-speed QNM shifts while leaving anisotropy-activated operators open. \\
QNM calibration & Direct timelike-scalar odd-sector QNM computations for stealth-Schwarzschild and Hayward backgrounds \cite{MukohyamaTakahashiTomikawaYingcharoenrat2023}. & A cleaned Hayward prior measure, admissible/proxy separation, and posterior-level pushforward into the dominant complex QNM plane. \\
Detector interpretation & Standard black-hole spectroscopy and next-generation detector forecasts \cite{BertiCardosoWill2006,MaggioreEtAl2020,AmaroSeoaneEtAl2017,BhagwatEtAl2022}. & Detector-whitened consistency modes comparing inherited, admissible anisotropic, and proxy anisotropic branches in the same covariance metric. \\
\hline\hline
\end{tabular}
\end{table}

\section{Cosmological input space and viability priors}
\label{sec:cosmoinput}

The transported object is the filtered support of a viable EFT branch, not an arbitrary modified-gravity theory. The posterior language introduced below defines the measure entering the map, while Sec.~\ref{subsec:benchmark_inputs} implements the compressed mock likelihood whose samples are pushed into ringdown space. This section also introduces the event-conditioned cosmological data vector used in the lift and identifies the null directions that survive cosmological constraints.

\subsection{Admissible cosmological parameter space}
\label{subsec:cosmoparamspace}

Let \(\mathfrak{C}\) denote a covariant theory class, for example a shift-symmetric Horndeski or quadratic-DHOST branch, together with the Jordan-frame choice in which matter is minimally coupled. The cosmological sector is then described by the Jordan-frame EFT data vector
\begin{equation*}
\Theta_{\rm DE}^{J}(t)
= \{H(t),\,M^2(t),\,u_a(t)\},
\end{equation*}
where \(u_a(t)\) denotes any basis spanning the surviving linear perturbation sector. Depending on the application, \(u_a\) may be taken to be the Bellini-Sawicki \(\alpha\) functions, a more directly covariant set of EFT coefficients, or a stability-built basis of the type used in recent Horndeski analyses \cite{BelliniSawicki2014,ZhengEtAl2025,StolznerEtAl2026}. Nothing below depends on a preferred basis; only the posterior measure carried by \(\Theta_{\rm DE}^{J}\) is used.

Given a cosmological data set \(d_{\rm cos}\), we define the raw cosmology-conditioned posterior measure by
\begin{equation}
P_{\rm cos}(\Theta_{\rm DE}^{J},\mathfrak{C}\,|\,d_{\rm cos})
\propto
\mathcal{L}_{\rm cos}\bigl(d_{\rm cos}\,|\,\Theta_{\rm DE}^{J},\mathfrak{C}\bigr)\,
\pi_{\rm raw}(\Theta_{\rm DE}^{J},\mathfrak{C}),
\label{eq:rawposterior}
\end{equation}
where \(\mathcal{L}_{\rm cos}\) is the cosmological likelihood and \(\pi_{\rm raw}\) is the pre-viability prior. The data set \(d_{\rm cos}\) may combine CMB anisotropies, BAO, supernovae, full-shape clustering, redshift-space distortions, and weak lensing. Such combinations already compress the viable EFT space enough to make a later strong-field pushforward informative \cite{ZhengEtAl2025,StolznerEtAl2026}. Equation~\eqref{eq:rawposterior} is the formal object that a full EFTCAMB/hi\_class analysis would provide; Sec.~\ref{subsec:benchmark_inputs} supplies a compressed mock posterior with the same structure: likelihood, priors, hard filters, posterior samples, and deterministic pushforward to ringdown variables.

\subsection{Hard viability filters}
\label{subsec:hardfilters}

The transport uses the filtered posterior obtained after imposing conservative theory and phenomenology cuts, rather than the raw posterior of Eq.~\eqref{eq:rawposterior},
\begin{align*}
P_{\rm surv}(\Theta_{\rm DE}^{J},\mathfrak{C}\,|\,d_{\rm cos})
&\propto
P_{\rm cos}(\Theta_{\rm DE}^{J},\mathfrak{C}\,|\,d_{\rm cos})\,
\Pi_{\rm hard},
\nonumber \\
\Pi_{\rm hard}
&\equiv
\Pi_{\rm bg}\,\Pi_T\,\Pi_{\rm stab}\,\Pi_{\rm deg}\,\Pi_{\rm loc}.
\end{align*}
Each factor in \(\Pi_{\rm hard}\) is an indicator functional that either accepts or rejects a branch of EFT histories. Some of these factors enter later sections as numerically implemented filters, while others act as structural admissibility requirements that define the class of branches to which the map is applied.

The factor \(\Pi_{\rm bg}\) enforces background admissibility: the EFT history must provide an acceptable homogeneous expansion history and a Jordan-frame description with standard matter geodesics. Because the map propagates already allowed cosmological information to the black-hole sector, this requirement is treated as a branch filter rather than as a separate phenomenological analysis.

The factor \(\Pi_T\) implements cosmological luminality in the low-redshift regime relevant to late-time EFT analyses. We write it schematically as
\begin{equation*}
\Pi_T
=
\Theta\!\left[
\epsilon_T-
\sup_{z\in\mathcal{Z}_*}
\bigl|c_T^2(z)-1\bigr|
\right],
\end{equation*}
where \(\mathcal{Z}_*\) is the redshift support of the events or source population and \(\epsilon_T\) is a small tolerance. In Horndeski-like bases this corresponds to imposing \(\alpha_T\simeq 0\) on FLRW \cite{EzquiagaZumalacarregui2017,CreminelliVernizzi2017}. In the odd-sector branch analyzed later, the numerical input is the low-redshift multimessenger bound from GW170817/GRB170817A, applied as a class assumption on the asymptotic homogeneous branch. For higher-redshift populations, this is a conditional extrapolation of the chosen EFT history rather than a model-independent statement. The filter \(\Pi_T\) acts only on the homogeneous tensor sector; it does not remove operators that vanish on FLRW and are activated only by anisotropic black-hole backgrounds.

The factor \(\Pi_{\rm stab}\) enforces linear stability along the cosmological branch. If \(\mathbf{K}\) is the kinetic matrix for the propagating perturbations and \(c_{s,I}^2\) the corresponding sound speeds, then the schematic requirement is
\begin{equation}
\lambda_I(\mathbf{K})>0,
\qquad
c_{s,I}^2>0,
\label{eq:stabreqs}
\end{equation}
throughout the redshift interval on which the EFT description is trusted. In practice one may supplement Eq.~\eqref{eq:stabreqs} with a bound against catastrophically fast tachyonic growth. Recent stable-basis analyses are particularly useful here because they implement these requirements directly in the sampling coordinates rather than only a posteriori \cite{StolznerEtAl2026}.

The factor \(\Pi_{\rm deg}\) imposes the class-specific degeneracy relations needed to remain within the chosen covariant theory orbit and to avoid unwanted Ostrogradsky modes. For DHOST-like theories these are the familiar relations among higher-derivative operators; for simpler Horndeski subclasses they are already built into the parameter basis. Operationally, we treat \(\Pi_{\rm deg}\) as a hard structural prior rather than as a separately sampled likelihood.

The factor \(\Pi_{\rm loc}\) encodes local-gravity admissibility: the surviving branch must recover GR in the early Universe or in sufficiently high-density environments, either through screening or through asymptotic suppression of the modified operators. Solar-System and binary-pulsar limits are not rederived in this work. Instead, \(\Pi_{\rm loc}\) is treated as a structural part of the class definition, so manifestly locally excluded branches never enter the cosmology-to-black-hole transport.

Together, these filters define a cosmologically viable branch and remove homogeneous-background failures before any strong-field extrapolation is attempted.

\subsection{Event-conditioned transport data}
\label{subsec:transportdata}

The lift depends not on the entire cosmological history, but on a finite amount of local information evaluated at the merger epoch of a given source. For an event at redshift \(z_*\) and cosmic time \(t_*\), with \(N\equiv \ln a\), we define the transport data vector
\begin{align}
\vartheta_{\rm cos}^{(M)}(t_*)
=
\Bigl\{
&H(t_*),\,\Omega_m(t_*),\,u_a(t_*),\,
\left.\frac{d u_a}{dN}\right|_{t_*},
\nonumber \\
&\ldots,\,
\left.\frac{d^M u_a}{dN^M}\right|_{t_*}
\Bigr\},
\label{eq:transportvector}
\end{align}
where the derivative order \(M\) is chosen to match the truncation order of the covariant finite jet introduced in Sec.~\ref{sec:framework}. Equation~\eqref{eq:transportvector} is the minimal cosmological information carried into the lift; all subsequent strong-field objects are functions of this vector together with the choice of theory class \(\mathfrak{C}\). In the fully worked inherited odd-sector branch of Sec.~\ref{sec:benchmarks} this truncation reduces to the minimal choice \(M=0\), because the exact rescaling law depends only on the asymptotic tensor-speed parameter itself rather than on time-derivative data.

The justification for this reduction is the enormous separation between cosmological and ringdown timescales. For both ground-based and space-based gravitational-wave observations, the remnant relaxes on a timescale that is negligible compared with \(H^{-1}(t_*)\). The cosmological EFT coefficients can therefore be treated as adiabatically frozen across the overlap shell
\begin{equation*}
r_g \ll r \ll H^{-1}(t_*),
\end{equation*}
so that \(\vartheta_{\rm cos}^{(M)}(t_*)\) acts as slowly varying boundary data for the black-hole EFT rather than as a dynamical input during the ringdown itself.

\subsection{Implemented compressed cosmological posterior and numerical inputs}
\label{subsec:benchmark_inputs}
\label{subsec:implemented_cosmo_posterior}

The ancillary script \posteriorcode\ implements the cosmological-posterior layer used in this analysis. The likelihood is compressed rather than Boltzmann-code-level, but it is sufficiently complete to exercise the full transport chain. The data vector is a deterministic, survey-calibrated simulation with fiducial background given by the Planck-2018 flat-$\Lambda$CDM point, $(\Omega_{m0},h,\sigma_8)=(0.315,0.674,0.811)$ \cite{Planck2018Parameters}; its distance/growth redshifts and fractional errors follow the structure of public BAO/RSD compressed summaries \cite{AlamBOSS2021}. The likelihood is transparent, repeatable, and readily replaceable by a full EFTCAMB/hi\_class chain. The compressed data vector contains BAO-like distance measurements, RSD-like growth measurements, and a GW170817-like homogeneous tensor-speed datum,
\begin{equation}
\bm d_{\rm mock}=\left\{D_M(z_i)/r_d,\,D_H(z_i)/r_d,\,f\sigma_8(z_j),\,\tau_T\right\},
\qquad
\tau_T\equiv 10^{15}\alpha_{T0},
\label{eq:mock_datavec}
\end{equation}
where $z_i=(0.38,0.51,0.61,0.85,1.48)$ and $z_j=(0.15,0.38,0.51,0.61,0.85)$. The fractional uncertainties mimic current-to-near-future compressed distance and growth information while remaining transparent and self-contained. The luminality prior is $\tau_T=0\pm1$, chosen near the conservative low-redshift multimessenger scale and applied only to the homogeneous FLRW tensor sector \cite{AbbottEtAl2017Multimessenger,EzquiagaZumalacarregui2017,CreminelliVernizzi2017}.

The EFT property functions in this executable example are
\begin{equation}
\alpha_I(a)=\alpha_{I0}\frac{\Omega_{\rm DE}(a)}{\Omega_{{\rm DE},0}},
\qquad I\in\{M,B,T\},
\label{eq:mock_alpha_param}
\end{equation}
with a fixed positive kinetic proxy. Distances use the flat-$\Lambda$CDM background. The growth sector uses the conservative quasistatic surrogate, normalized to the usual GR growth-index approximation \cite{Linder2005Growth},
\begin{align}
f(a)&=\Omega_m(a)^{\gamma(a)}\left[1+0.45\alpha_B(a)-0.25\alpha_M(a)\right],
\nonumber\\
\gamma(a)&=0.545-0.15\alpha_M(a)+0.08\alpha_B(a),
\label{eq:growth_surrogate}
\end{align}
and $D(a)/D(1)=\exp[-\int_a^1 f\,d\ln a]$. We sample $(\Omega_{m0},h,\sigma_8,\alpha_{M0},\alpha_{B0},\tau_T)$ with an adaptive independent importance sampler. The hard cuts require a viable background box, a positive scalar kinetic proxy $Q_s=1+1.5\alpha_B(a)^2$, a conservative sound-speed proxy $c_s^2=1+0.40\alpha_B(a)-0.20\alpha_M(a)>0.05$ on $0\le z\le3$, and the homogeneous tensor-speed window encoded in $\tau_T$. The script writes the mock data, posterior samples, summary JSON, and figures used below; no external downloads are required to regenerate this layer.

\begin{figure}[t]
\centering
\subfloat[Marginalized posterior slices entering the finite-jet lift.\label{fig:cosmo_posterior_slices}]{%
  \includegraphics[width=0.62\textwidth]{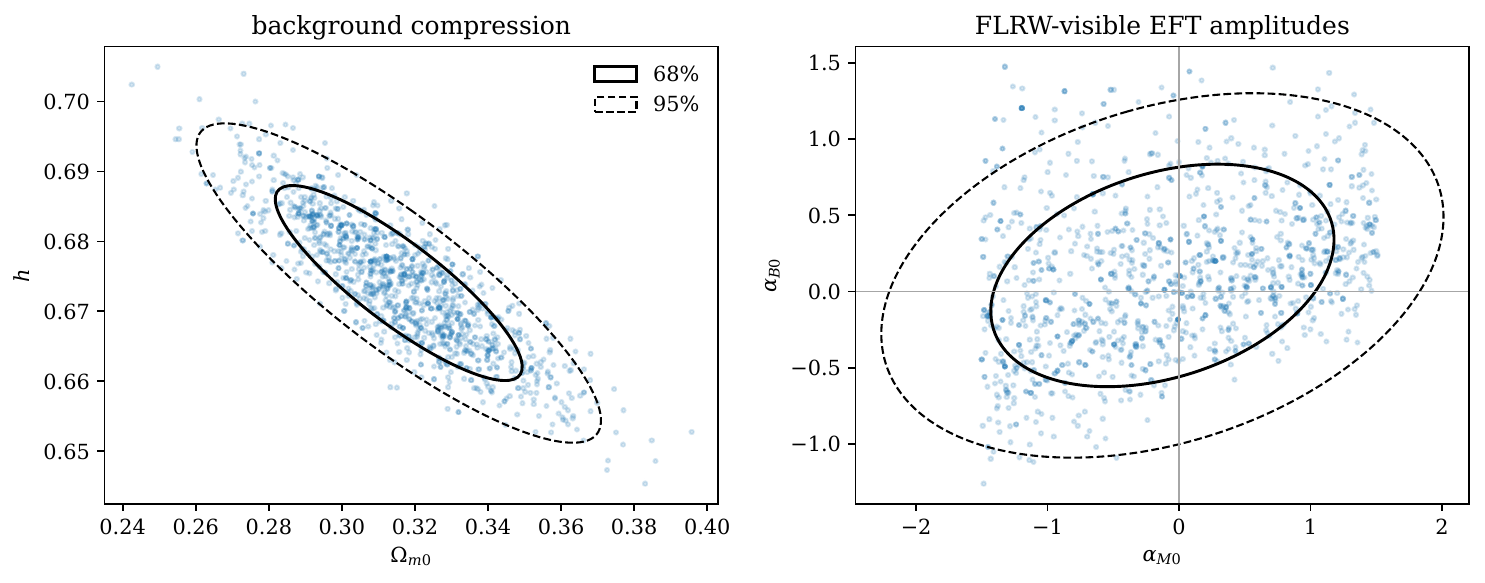}}
\hfill
\subfloat[Posterior pushforward into the dominant odd-sector plane.\label{fig:posterior_pushforward}]{%
  \includegraphics[width=0.34\textwidth]{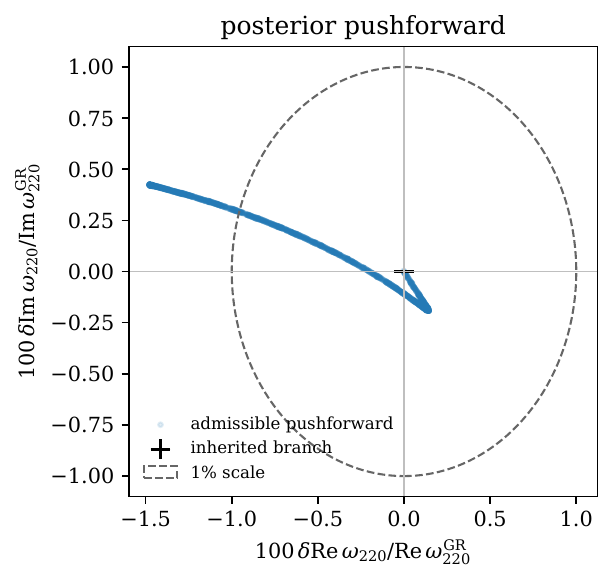}}
\caption{Implemented cosmological posterior and ringdown pushforward. Panel (a) shows the sampled compressed posterior for the background and EFT amplitudes. Panel (b) pushes the same samples to the dominant odd-sector QNM plane. The inherited branch is visually indistinguishable from the origin at this scale because the tensor-speed posterior gives $|\delta\omega_{220}/\omega_{220}|\sim10^{-15}$, while the FLRW-silent anisotropy branch occupies an admissible subpercent region.}
\label{fig:cosmo_posterior_pushforward}
\end{figure}

Table~\ref{tab:posterior_summary} gives the posterior summary used in the rest of the paper. The qualitative outcome does not depend on the detailed mock realization: the inherited homogeneous tensor-speed amplitude is driven to the $10^{-15}$ level, whereas the FLRW-silent anisotropy variable is controlled by its structural prior and by the black-hole-side stability interval. Figure~\ref{fig:cosmo_posterior_pushforward} shows the same result visually.

\begin{table}[H]
\caption{Posterior summary from the executable compressed cosmological likelihood in \posteriorcode. The data are simulated, but the likelihood, priors, stability cuts, posterior samples, and ringdown pushforward are all implemented in the ancillary workflow. The anisotropy rows are derived quantities obtained by pushing the accepted samples through the Hayward-calibrated odd-sector response.}
\label{tab:posterior_summary}
\centering
\begin{tabular}{lll}
\hline\hline
Quantity & Posterior summary & Role \\
\hline
$\Omega_{m0}$ & $0.3145^{+0.0231}_{-0.0211}$ & background \\
$h$ & $0.6741^{+0.0092}_{-0.0090}$ & background \\
$\sigma_8$ & $0.7970^{+0.0663}_{-0.0667}$ & growth \\
$\alpha_{M0}$ & $-0.171^{+1.07}_{-0.942}$ & FLRW-visible EFT \\
$\alpha_{B0}$ & $0.098^{+0.478}_{-0.464}$ & FLRW-visible EFT \\
$\tau_T=10^{15}\alpha_{T0}$ & $-0.0281^{+1.01}_{-0.997}$ & homogeneous tensor speed \\
$|\delta q_{220}^{\rm inh}|_{95}$ & $2.9e-15$ & inherited ringdown \\
$\eta_{\rm QNM,95}$ & $0.0216$ & FLRW-silent prior \\
$|\delta_R|_{95}$ & $0.0146$ & Hayward odd response \\
$|\delta_I|_{95}$ & $0.00419$ & Hayward odd response \\
\hline\hline
\end{tabular}

\end{table}

For comparison with the analytic and literature-calibrated sections below, Table~\ref{tab:cosmo_inputs} also records the reduced priors used when an expression is intentionally evaluated outside the posterior script. The proxy Hayward row is included only for detector-reach calibration and is not promoted to a cosmology-demonstrated posterior.

\begin{table}[H]
\caption{Reduced cosmological and strong-field inputs used outside the executable posterior script. The first row is also implemented in the posterior through $\tau_T=10^{15}\alpha_{T0}$. The proxy Hayward continuation is retained only as a detector-reach diagnostic.}
\label{tab:cosmo_inputs}
\small
\begin{tabular*}{\textwidth}{@{\extracolsep{\fill}}llll@{}}
\hline\hline
\parbox[t]{0.20\textwidth}{Quantity} & \parbox[t]{0.14\textwidth}{Mean / support} & \parbox[t]{0.18\textwidth}{Width or covariance} & \parbox[t]{0.36\textwidth}{Role in this work} \\
\hline
\parbox[t]{0.20\textwidth}{Inherited tensor-speed parameter $\alpha_T$} & \parbox[t]{0.14\textwidth}{0} & \parbox[t]{0.18\textwidth}{$|\alpha_T|\lesssim 6\times10^{-15}$} & \parbox[t]{0.36\textwidth}{Multimessenger luminality prior used in Eqs.~\eqref{eq:alphaT_bound_explicit} and \eqref{eq:inherited_bound_final} \cite{AbbottEtAl2017Multimessenger}} \\
\parbox[t]{0.20\textwidth}{Toy lift vector $(y_1,y_2)$} & \parbox[t]{0.14\textwidth}{$(0,0.03)$} & \parbox[t]{0.18\textwidth}{$\mathbf C_y=10^{-4}\mathbf I_2$} & \parbox[t]{0.36\textwidth}{Nontrivial $N=1$ lift validation with one FLRW null direction} \\
\parbox[t]{0.20\textwidth}{Admissible Hayward interval} & \parbox[t]{0.14\textwidth}{$0<\eta_{\rm QNM}<2.17\times10^{-2}$} & \parbox[t]{0.18\textwidth}{$\sigma_\eta=\eta_{\rm max}/2$} & \parbox[t]{0.36\textwidth}{Positive stable asymptotically luminal interval used in the admissible anisotropy prior} \\
\parbox[t]{0.20\textwidth}{Proxy Hayward continuation} & \parbox[t]{0.14\textwidth}{centered at 0} & \parbox[t]{0.18\textwidth}{$\sigma_\eta=3.0\times10^{-2}$} & \parbox[t]{0.36\textwidth}{Literature-calibrated reach estimate only; not cosmology demonstrated} \\
\hline\hline
\end{tabular*}
\end{table}

\subsection{Cosmological projection and null directions}
\label{subsec:nullspace}

Even a sharply constrained cosmological posterior does not determine all directions in the parent theory space. Linearizing around a viable cosmological branch, let \(\bm{\theta}_{\rm parent}\) denote a finite set of parent-theory parameters and let \(\mathbf{P}_{\rm FLRW}\) be the Jacobian that maps them to cosmological observables. Then
\begin{equation*}
\delta d_{\rm cos}
=
\mathbf{P}_{\rm FLRW}\,\delta\bm{\theta}_{\rm parent}.
\end{equation*}
The parent perturbation naturally decomposes as
\begin{equation*}
\delta\bm{\theta}_{\rm parent}
=
\delta\bm{\theta}_{\parallel}
+
\delta\bm{\theta}_{\perp},
\qquad
\delta\bm{\theta}_{\perp}\in \ker \mathbf{P}_{\rm FLRW}.
\end{equation*}
The component \(\delta\bm{\theta}_{\parallel}\) is seen by cosmology. The component \(\delta\bm{\theta}_{\perp}\) is symmetry silent on FLRW, but it need not be irrelevant near a black hole. This null space is the mathematical reason that a cosmological EFT posterior cannot be naively read as a black-hole EFT posterior.

We therefore do not set \(\delta\bm{\theta}_{\perp}=0\) by hand, nor do we treat it as completely arbitrary. Instead, Sec.~\ref{sec:framework} defines a covariant lift that regularizes these null directions within a chosen theory class. The filtered cosmological posterior is then pushed through that lift into ringdown observable space.

\section{Theoretical framework}
\label{sec:framework}

This section puts cosmological constraints and black-hole ringdown observables in a common EFT language. The obstacle is precisely the one identified above: cosmology constrains the isotropic FLRW projection, while black-hole perturbations can probe anisotropy-activated operators that vanish on FLRW\@. The relevant theory space is therefore wider than the cosmological EFT coefficients alone; it is the covariant parent theory whose different background projections describe the cosmological and strong-field regimes \cite{Gubitosi2013,Bloomfield2013,BelliniSawicki2014,FrusciantePerenon2020,MukohyamaYingcharoenrat2022,MukohyamaSerailleTakahashiYingcharoenrat2025}.

The construction starts from the Jordan-frame cosmological EFT, lifts it to a finite covariant parent description, embeds the result in the arbitrary-background EFT for timelike scalar profiles, and finally projects to the static, spherically symmetric black-hole background relevant for ringdown. The next section turns these steps into an explicit response from cosmological data to black-hole observables.

\subsection{Parent theory space and timelike scalar regime}
\label{subsec:parent}

We work in scalar-tensor theories containing a single additional scalar field \(\phi\) with timelike gradient,
\begin{equation*}
X \equiv g^{\mu\nu}\nabla_{\mu}\phi\,\nabla_{\nu}\phi < 0,
\end{equation*}
throughout the spacetime region connecting the asymptotic cosmological domain to the black-hole near zone. This is the regime in which the EFT of perturbations with a timelike scalar profile is applicable on arbitrary backgrounds \cite{MukohyamaYingcharoenrat2022}. For definiteness we imagine that the underlying covariant action lies in a class that is closed under invertible conformal/disformal transformations, such as shift-symmetric Horndeski or quadratic DHOST, although much of the discussion is basis independent \cite{MukohyamaTakahashiYingcharoenrat2022,MukohyamaSerailleTakahashiYingcharoenrat2025}.

A convenient abstract representation is
\begin{equation*}
S_{\rm parent}
= \int \dd^4x\,\sqrt{-g}\,\mathcal{L}_{\rm parent}[g_{\mu\nu},\phi]
+ S_m[g_{\mu\nu},\psi_m],
\end{equation*}
with
\begin{equation}
\mathcal{L}_{\rm parent}
= \sum_A \mathcal{F}_A(\phi,X)\,\mathcal{O}_A[g_{\mu\nu},\phi].
\label{eq:Lparent_expansion}
\end{equation}
In the shift-symmetric case the coefficient functions reduce to \(\mathcal{F}_A(X)\), which will simplify the lift problem below. Equation~\eqref{eq:Lparent_expansion} is not tied to a particular covariant basis; it simply makes explicit that the cosmological EFT coefficients are projections of a more informative parent data set.

\subsection{Cosmological EFT in unitary gauge}
\label{subsec:deeft}

On a spatially flat FLRW background and in unitary gauge, \(\phi = \phi_0(t)\), the EFT of dark energy in the Jordan frame can be written as \cite{Gubitosi2013,Bloomfield2013,FrusciantePerenon2020}
\begin{align}
S_{\rm DE}^{(J)}
= \int \dd^4x\,\sqrt{-g}\Bigg\{&
\frac{M^2(t)}{2}R - \Lambda(t) - c(t)g^{00}
\nonumber\\
&+ \frac{M_2^4(t)}{2}(\delta g^{00})^2
- \frac{\bar M_1^3(t)}{2}\,\delta g^{00}\,\delta K
\nonumber\\
&- \frac{\bar M_2^2(t)}{2}(\delta K)^2
\nonumber\\
&- \frac{\bar M_3^2(t)}{2}\,\delta K^{\mu}{}_{\nu}\,\delta K^{\nu}{}_{\mu}
\nonumber\\
&+ \frac{\hat M^2(t)}{2}\,\delta g^{00}\,\delta {}^{(3)}\!R
+ \cdots\Bigg\}
\nonumber\\
&+ S_m[g_{\mu\nu},\psi_m].
\label{eq:EFTDE}
\end{align}
Here \(\delta K \equiv K-3H\) and \({}^{(3)}\!R\) is the intrinsic Ricci scalar of the constant-time hypersurfaces. Equation~\eqref{eq:EFTDE} is the background and linear-perturbation EFT relevant for large-scale structure and gravitational-wave propagation on cosmological backgrounds.

For practical purposes it is convenient to collect the cosmological information into a reduced Jordan-frame data vector
\begin{equation*}
\Theta_{\rm DE}^{J}(t)
= \{ H(t),\,M^2(t),\,u_a(t)\},
\end{equation*}
where \(u_a(t)\) denotes any reduced basis appropriate for the chosen parent class. In Horndeski-like applications one may trade the coefficients in Eq.~\eqref{eq:EFTDE} for the Bellini-Sawicki property functions \cite{BelliniSawicki2014}. Following GW170817 and GRB170817A, viable cosmological models are typically restricted to a luminal tensor speed on FLRW, \(\alpha_T \simeq 0\), together with the usual ghost and gradient-stability conditions \cite{EzquiagaZumalacarregui2017,CreminelliVernizzi2017}. We keep the notation \(u_a\) basis agnostic because the black-hole projection developed below only requires that \(\Theta_{\rm DE}^{J}\) faithfully parametrizes the surviving cosmological EFT coefficients.

\subsection{Controlled covariant lift}
\label{subsec:lift}

The cosmological EFT coefficients are not yet the parent theory; they are its FLRW shadow. Schematically,
\begin{equation*}
u_a(t)
= U_a\!\left[\mathcal{F}_A,\partial_X\mathcal{F}_A,\partial_X^2\mathcal{F}_A,\ldots;\,\vev X(t),H(t),\dot H(t),\ldots\right],
\end{equation*}
where \(\vev X(t)\) is the homogeneous background value and the maps \(U_a\) follow from the covariant-to-EFT dictionary of the chosen scalar-tensor class. Because FLRW is homogeneous and isotropic, any coefficient multiplying traceless background tensors is projected out. This implies that the inverse problem is underdetermined: cosmology constrains only the isotropic projection of the parent theory \cite{MukohyamaYingcharoenrat2022,MukohyamaSerailleTakahashiYingcharoenrat2025}.

We make the inverse problem finite by reconstructing only a local jet of the parent functions around the asymptotic cosmological value
\begin{equation*}
X_{\infty} \equiv \vev X(t_*),
\end{equation*}
evaluated at the merger epoch \(t_*\). In the shift-symmetric case we write
\begin{equation}
\mathcal{F}_A(X)
= \sum_{n=0}^{N} \frac{f_{A,n}}{n!}\,(X-X_{\infty})^n,
\label{eq:finite_jet}
\end{equation}
and determine the coefficients \(f_{A,n}\) by matching not only the values of \(u_a\) at \(t_*\) but also a finite number of time derivatives. The lift is defined as the regularized pseudo-inverse
\begin{equation*}
J^N\mathcal{F}
= \arg\min_{\{f_{A,n}\}}\left[\chi^2_{\rm cos} + \lambda_{\rm reg}\,\mathcal{R}(\{f_{A,n}\})\right],
\end{equation*}
subject to the hard constraints implied by degeneracy, stability, and the cosmological luminality prior. The functional \(\mathcal{R}\) suppresses motion along the null directions that are invisible on FLRW\@. Operationally, the lift is implemented by choosing a finite jet order \(N\), evaluating the Jacobian of the FLRW projection on a fiducial viable branch, projecting to the admissible singular-vector subspace selected by \(\Pi_{\rm hard}\), and then computing the regularized inverse in that reduced space. In the explicit inherited odd-sector branch of Sec.~\ref{sec:benchmarks} the lift truncates already at \(N=0\), so regularization does not generate a nontrivial null-space ambiguity inside the worked example itself. This lift is a regularized right-inverse of the cosmological projection. It replaces the stronger, and generally unjustified, assumption that FLRW EFT data uniquely determine the strong-field theory.

\subsection{Arbitrary-background EFT with timelike scalar profile}
\label{subsec:arbbackground}

Once a parent jet has been reconstructed, it can be projected onto the arbitrary-background EFT appropriate for a timelike scalar profile \cite{MukohyamaYingcharoenrat2022}. In unitary gauge the scalar defines a preferred foliation with unit normal and induced metric
\begin{equation*}
n_{\mu} = -\frac{\nabla_{\mu}\phi}{\sqrt{-X}},
\qquad
h_{\mu\nu} = g_{\mu\nu}+n_{\mu}n_{\nu},
\end{equation*}
and extrinsic curvature
\begin{equation*}
K_{\mu\nu} = h_{\mu}{}^{\alpha}\nabla_{\alpha}n_{\nu}.
\end{equation*}
The intrinsic curvature of the constant-\(\phi\) hypersurfaces is denoted by \({}^{(3)}\!R_{\mu\nu}\). On a generic background it is useful to separate trace and traceless pieces,
\begin{equation*}
\vev\sigma_{\mu\nu}
\equiv \vev K_{\mu\nu} - \frac{1}{3}\vev K\,h_{\mu\nu},
\qquad
\vev r_{\mu\nu}
\equiv {}^{(3)}\!\vev R_{\mu\nu} - \frac{1}{3}{}^{(3)}\!\vev R\,h_{\mu\nu},
\end{equation*}
which vanish on FLRW but not on a black-hole background.

The arbitrary-background EFT may be organized as
\begin{equation*}
S_{\rm arb}
= \int \dd^4x\,\sqrt{-g}\,\bigl(\mathcal{L}_{\rm tad}+\mathcal{L}_{\rm quad}+\cdots\bigr),
\end{equation*}
where the quadratic sector contains operators built from the covariant fluctuations of the lapse, extrinsic curvature, and intrinsic curvature of the preferred foliation. We denote the full coefficient set by
\begin{equation*}
\mathcal{C}_I^{J}(x)
= \{f,\Lambda,c,m_4^2,M_2^4,M_3^3,M_6,\lambda_I{}^{\mu}{}_{\nu},\mu_I,\ldots\},
\end{equation*}
with the understanding that residual spatial diffeomorphisms impose nontrivial consistency relations among these coefficients \cite{MukohyamaYingcharoenrat2022}. At the schematic level the quadratic action includes terms such as
\begin{align*}
\mathcal{L}_{\rm quad} \supset {}&
\frac{M_2^4}{2}(\delta N)^2
- \frac{M_3^3}{2}\,\delta N\,\delta K
\nonumber\\
&- m_4^2\left(\delta K^2 - \delta K^{\mu}{}_{\nu}\delta K^{\nu}{}_{\mu}\right)
\nonumber\\
&+ M_6\,\vev\sigma^{\mu}{}_{\nu}\,\delta K^{\nu}{}_{\alpha}\delta K^{\alpha}{}_{\mu}
+ \cdots.
\end{align*}
The last operator is particularly important for our purposes. Since \(\vev\sigma^{\mu}{}_{\nu}=0\) on FLRW, it is invisible to homogeneous cosmology and to the usual cosmological tensor-speed bound, but it becomes active on anisotropic black-hole backgrounds and can modify odd-parity propagation in the near zone \cite{MukohyamaSerailleTakahashiYingcharoenrat2025}.

The lifted parent theory determines the arbitrary-background EFT coefficients through a covariant dictionary,
\begin{equation*}
\mathcal{C}_I^{J}(x)
= \mathcal{D}_I\!\left[J^N\mathcal{F};g_{\mu\nu},\phi\right],
\end{equation*}
which is known explicitly for broad scalar-tensor classes and can be specialized to particular operator bases when needed \cite{MukohyamaYingcharoenrat2022,MukohyamaTakahashiYingcharoenrat2022}.

\subsection{Frame transformations and almost Einstein frame}
\label{subsec:frames}

Cosmological constraints are typically reported in the Jordan frame, where matter is minimally coupled. For black-hole perturbations it is often convenient to work in a frame where the coefficient of the Ricci scalar is constant. We therefore allow an invertible conformal/disformal transformation,
\begin{equation*}
g^{E}_{\mu\nu}
= C(\phi,X)\,g^{J}_{\mu\nu} + D(\phi,X)\,\nabla_{\mu}\phi\nabla_{\nu}\phi,
\end{equation*}
and choose \(C\) such that the transformed Einstein-Hilbert term is canonical,
\begin{equation*}
f^{E} = M_{\rm Pl}^2.
\end{equation*}
The transformed EFT coefficients are denoted by
\begin{equation*}
\mathcal{C}_I^{E}(x) = \mathcal{T}_I\!\left[\mathcal{C}_J^{J}(x);C,D\right].
\end{equation*}
The arbitrary-background conformal/disformal transport has been worked out explicitly in the timelike-scalar EFT, and the transformation preserves the consistency relations among EFT coefficients when the map is invertible \cite{MukohyamaSerailleTakahashiYingcharoenrat2025}.

\subsection{Static, spherically symmetric black holes and parity sectors}
\label{subsec:sphericalbh}

To connect with ringdown we specialize to a static, spherically symmetric background,
\begin{equation*}
\dd s^2 = -A(r)\dd t^2 + \frac{\dd r^2}{B(r)} + r^2\dd\Omega^2,
\qquad
\phi(t,r) = q t + \psi(r),
\end{equation*}
with \(X(r)<0\) in the domain of interest. The timelike component \(q\neq 0\) allows the preferred foliation to remain regular across the horizon within the class of solutions described by the timelike-scalar EFT \cite{MukohyamaYingcharoenrat2022}. The black-hole regime and the cosmological regime overlap in the shell
\begin{equation}
r_g \ll r \ll H^{-1}(t_*),
\label{eq:matching_shell}
\end{equation}
where cosmological evolution is adiabatically slow on the ringdown timescale and the asymptotic EFT coefficients may be frozen at the merger epoch.

On the constant-\(\phi\) hypersurfaces we introduce the radial unit vector \(s^{\mu}\) and the projector onto the two-sphere,
\begin{equation*}
\Pi^{\mu}{}_{\nu} = h^{\mu}{}_{\nu} - s^{\mu}s_{\nu}.
\end{equation*}
A generic spatial rank-two coefficient then decomposes as
\begin{align*}
T^{\mu}{}_{\nu}
&= T_r(r)\,s^{\mu}s_{\nu} + T_{\Omega}(r)\,\Pi^{\mu}{}_{\nu}
\nonumber\\
&= T_0(r)\,h^{\mu}{}_{\nu}
+ T_1(r)\left(s^{\mu}s_{\nu}-\frac{1}{2}\Pi^{\mu}{}_{\nu}\right).
\end{align*}
This decomposition cleanly separates isotropic and anisotropic structures on the three-slice. It also motivates the classification of EFT coefficients into inherited coefficients, induced anisotropic coefficients that vanish on FLRW but are fixed by the parent jet, and genuinely strong-field coefficients that remain free at the chosen truncation order.

At the level of perturbations, spherical harmonics split the dynamics into odd and even sectors. The odd sector is governed by a generalized Regge-Wheeler equation derived within the timelike-scalar EFT \cite{MukohyamaTakahashiYingcharoenrat2022,MukohyamaTakahashiTomikawaYingcharoenrat2023},
\begin{equation*}
\frac{\dd^2\Psi^{\rm odd}_{\ell m}}{\dd r_*^2}
+ \left[\omega^2 - V_{\rm odd}\bigl(r;\Theta_{\rm odd}(r)\bigr)\right]\Psi^{\rm odd}_{\ell m}=0,
\end{equation*}
where \(r_*\) is the tortoise coordinate. The even sector is a coupled metric-scalar system,
\begin{equation*}
\mathbf{D}_{\rm even}\!\left(r,\omega;\Theta_{\rm even}(r)\right)\mathbf{u}_{\ell m}(r)=0,
\end{equation*}
and recent work has begun to isolate tractable subsectors by adding the scordatura regulator to avoid strong coupling around stealth solutions \cite{MukohyamaTakahashiTomikawaYingcharoenrat2025}.

\subsection{Formal consistency map}
\label{subsec:formal_map}

The ingredients above define the composite map
\begin{equation}
\Theta_{\rm BH}^{E}(r;t_*)
= \Pi_{\rm BH}\circ \mathcal{T}_{J\to E}\circ \mathcal{D}_{\rm arb}\circ \mathcal{L}_N\bigl[\Theta_{\rm DE}^{J}(t_*)\bigr],
\label{eq:master_map}
\end{equation}
where \(\mathcal{L}_N\) is the controlled finite-jet lift, \(\mathcal{D}_{\rm arb}\) the covariant dictionary to the arbitrary-background EFT, \(\mathcal{T}_{J\to E}\) the frame transformation, and \(\Pi_{\rm BH}\) the projection onto the static, spherically symmetric black-hole background and its parity sectors. The next section turns Eq.~\eqref{eq:master_map} into a response from cosmological posteriors to black-hole ringdown observables.

\section{Constructing the cosmology-to-black-hole EFT map}
\label{sec:map}

Section~\ref{sec:framework} introduced the projections and frame transformations linking cosmological modified gravity to black-hole perturbations. We expand around a fiducial cosmologically viable solution, lift the cosmological data to a finite parent-theory jet, transport that jet to the black-hole near zone, and project the result onto parity-resolved perturbation operators and quasinormal-mode (QNM) observables. This gives a response matrix that carries cosmological covariance directly to ringdown parameters.

\subsection{Cosmological input vector and finite-jet response}
\label{subsec:data_vector}

Let \(t_*\) denote the merger epoch of the black-hole event whose ringdown is to be modeled. Instead of using only the instantaneous cosmological EFT coefficients, we define an augmented data vector,
\begin{equation}
\bm{y}_* \equiv \bigl\{u_a(t_*),\;H_*^{-1}\dot u_a(t_*),\;\ldots,\;H_*^{-M}\partial_t^{M}u_a(t_*)\bigr\},
\label{eq:yvec}
\end{equation}
where \(H_*\equiv H(t_*)\) renders each entry dimensionless. The inclusion of a finite number of time derivatives is important: two covariant theories can agree on the instantaneous FLRW EFT coefficients yet differ in the first few derivatives, and these derivative directions are precisely what the finite-jet lift is designed to resolve.

Expanding the parent functions as in Eq.~\eqref{eq:finite_jet}, we collect the jet coefficients into a vector
\begin{equation*}
\bm{f} \equiv \{f_{A,n}\},
\qquad n=0,\ldots,N.
\end{equation*}
Linearizing the cosmological projection around a fiducial cosmologically viable model gives
\begin{equation}
\delta \bm{y}_* = \mathbf{J}\,\delta \bm{f},
\qquad
J_{i\alpha} \equiv \left.\frac{\partial y_i}{\partial f_{\alpha}}\right|_{\rm fid},
\label{eq:Jmatrix}
\end{equation}
where \(\alpha\equiv(A,n)\) is a composite jet index. We treat the regularizer as a Gaussian prior precision on the finite jet, not as a post-processing penalty on a frequentist estimator. Thus
\begin{equation}
p(\delta\bm f|\delta\hat{\bm y}_*)\propto
\exp\left[-\frac12(\delta\hat{\bm y}_*-\mathbf J\delta\bm f)^{\mathsf T}\mathbf C_y^{-1}(\delta\hat{\bm y}_*-\mathbf J\delta\bm f)
-\frac12\delta\bm f^{\mathsf T}\mathbf\Lambda_f\delta\bm f\right],
\qquad
\mathbf\Lambda_f\equiv\lambda_{\rm reg}\mathbf R+\mathbf P_{\rm deg},
\label{eq:lift_posterior_density}
\end{equation}
where \(\mathbf{R}\) penalizes directions invisible on FLRW and \(\mathbf P_{\rm deg}\) denotes hard or stiff quadratic degeneracy constraints. The posterior covariance and mean are
\begin{align}
\mathbf C_f
&=\left(\mathbf J^{\mathsf T}\mathbf C_y^{-1}\mathbf J+\mathbf\Lambda_f\right)^{-1},
\label{eq:Cf}
\\
\widehat{\delta\bm f}
&=\mathbf C_f\mathbf J^{\mathsf T}\mathbf C_y^{-1}\delta\hat{\bm y}_*
\equiv \mathbf L\,\delta\hat{\bm y}_*,
\qquad
\mathbf L\equiv\mathbf C_f\mathbf J^{\mathsf T}\mathbf C_y^{-1}.
\label{eq:Lmatrix}
\end{align}
Equations~\eqref{eq:Jmatrix}--\eqref{eq:Lmatrix} provide a concrete implementation of the controlled lift introduced abstractly in Sec.~\ref{sec:framework}. This convention is the one used in the numerical lift and in the executable posterior script. If one instead interprets the same algebra as a deterministic Tikhonov estimator, the sampling covariance would be \(\mathbf L\mathbf C_y\mathbf L^{\mathsf T}\); that is a different object and is not the covariance propagated in this work. When \(\mathbf{J}^{\mathsf T}\mathbf{C}_y^{-1}\mathbf{J}\) is numerically ill-conditioned, the implementation should be understood as an SVD-truncated inverse on the admissible singular-vector subspace, with \(\mathbf{R}\) penalizing the remaining FLRW-silent directions rather than attempting to invert them exactly.

\subsection{A nontrivial \texorpdfstring{$N=1$}{N=1} lift demonstration}
\label{subsec:lift_demo}

To test the controlled lift in the presence of a genuine FLRW null direction, consider the smallest nontrivial toy jet,
\(\delta \bm f=(\delta f_0,\delta f_1,\delta m_0)^{\mathsf T}\),
for which the cosmological summary vector
\(\delta \bm y_*=(\delta y_1,\delta y_2)^{\mathsf T}\)
obeys
\begin{equation}
\delta \bm y_* =
\begin{pmatrix}
1 & 0 & 0 \\
0 & 1 & 1
\end{pmatrix}
\delta \bm f,
\label{eq:toy_lift_matrix}
\end{equation}
with
\(\mathbf C_y={\rm diag}(\sigma_y^2,\sigma_y^2)\)
and
\(\sigma_y=10^{-2}\).
The first datum fixes the inherited direction \(\delta f_0\), whereas the second is sensitive only to the image combination \(\delta f_1+\delta m_0\) and leaves the orthogonal completion direction unresolved. For the illustrative datum
\(\delta\hat{\bm y}_*=(0,0.03)^{\mathsf T}\)
we regularize the null sector with
\begin{equation}
\mathbf R_{\rm toy}
=
{\rm diag}\!\left(0,\sigma_1^{-2},9\sigma_1^{-2}\right),
\qquad
\sigma_1=10^{-2},
\label{eq:toy_lift_reg}
\end{equation}
so that the image direction remains lightly constrained while the completion coefficient \(\delta m_0\) is damped more strongly than \(\delta f_1\).

\begin{figure}[t]
\centering
\subfloat[Null stripe and regularized posteriors in the \((\delta f_1,\delta m_0)\) plane.\label{fig:lift_ellipses}]{%
  \includegraphics[width=0.485\textwidth]{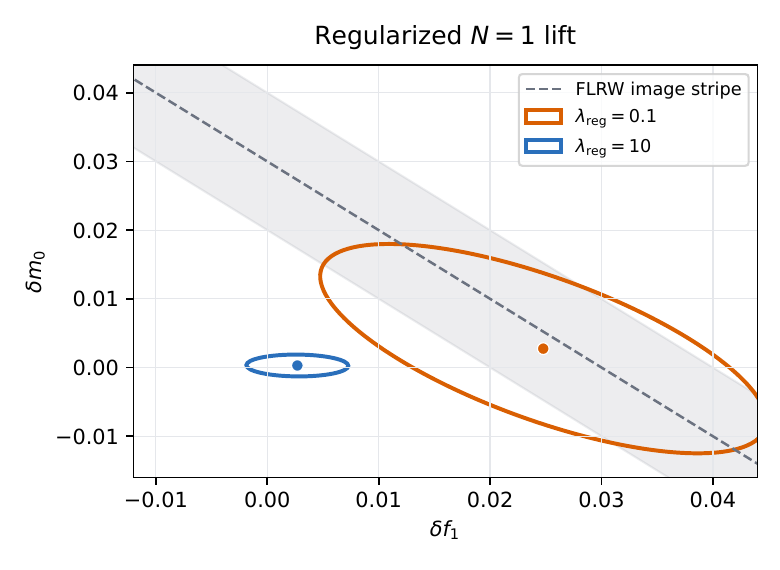}}
\hfill
\subfloat[Near-horizon completion coefficient induced by Eq.~\eqref{eq:toy_M6_predictor}.\label{fig:lift_m6}]{%
  \includegraphics[width=0.485\textwidth]{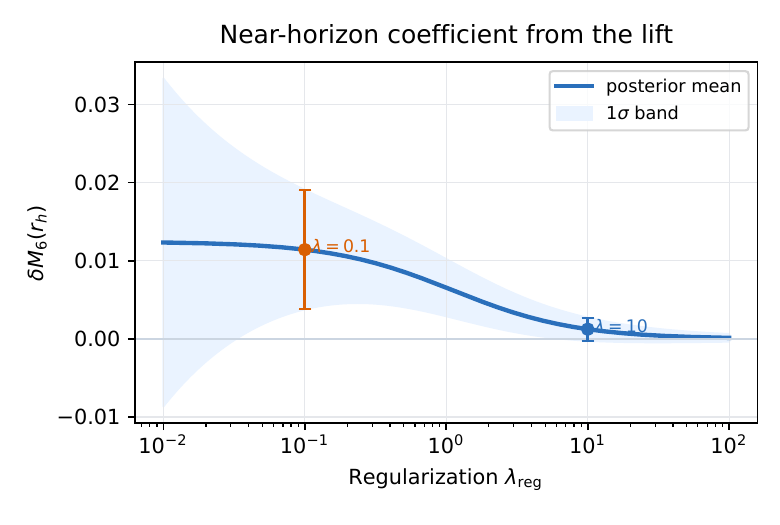}}
\caption{Nontrivial validation of the controlled finite-jet lift. Panel~(a) shows the exact FLRW image-compatible null stripe together with the 68\% posteriors obtained after regularization. Panel~(b) propagates the same posteriors to a near-horizon anisotropy coefficient. The exercise does not claim a fundamental parent theory for the toy model; its purpose is to show that the lift behaves controllably when the cosmological data leave an actual null direction unresolved.}
\label{fig:lift_demo}
\end{figure}

To turn the lifted jet into a black-hole-side prediction we define a near-horizon proxy coefficient,
\begin{equation}
\delta M_6(r_h)=\delta m_0+0.35\,\delta f_1,
\label{eq:toy_M6_predictor}
\end{equation}
which mimics an anisotropy-activated completion sourced partly by a first-derivative jet coefficient and partly by an FLRW-silent background term. Figure~\ref{fig:lift_demo} shows the resulting posterior geometry. Panel~(a) displays the exact \(\lambda_{\rm reg}=0\) image-compatible null stripe together with the 68\% posteriors for \(\lambda_{\rm reg}=0.1\) and \(\lambda_{\rm reg}=10\). Panel~(b) propagates the same posteriors to the near-horizon predictor of Eq.~\eqref{eq:toy_M6_predictor}. For \(\lambda_{\rm reg}=0.1\) the lift gives
\(\delta M_6(r_h)=(1.14\pm0.76)\times10^{-2}\),
whereas for \(\lambda_{\rm reg}=10\) it contracts to
\(\delta M_6(r_h)=(1.25\pm1.46)\times10^{-3}\).
Across
\(10^{-2}\le\lambda_{\rm reg}\le10^{2}\),
the posterior mean and variance drift smoothly. The toy predictor is not fundamental; it shows that, in an underdetermined \(N>0\) problem, the cosmology image, FLRW null stripe, and induced near-horizon coefficient can be tracked explicitly as the regularization is varied.

The toy variables also make the connection to the physical branch explicit. The coefficient \(\delta f_1\) represents a cosmology-image jet direction, while \(\delta m_0\) represents the local linearization of an FLRW-silent anisotropic completion such as the coefficient multiplying the \(M_6\bar\sigma^\mu{}_{\nu}\delta K^\nu{}_{\alpha}\delta K^\alpha{}_{\mu}\) operator. The predictor in Eq.~\eqref{eq:toy_M6_predictor} should therefore be read as a controlled local model of how a null cosmological direction can reappear as a near-horizon odd-sector coefficient. It is not used to replace the Hayward QNM calibration of Sec.~\ref{subsec:lit_calibrated_odd}; it validates the inverse step that makes such anisotropic completions transportable.

\subsection{Background transport from the overlap region to the near zone}
\label{subsec:transport_bg}

The lift must then be evaluated on the black-hole background. Let
\begin{equation*}
\bm{Y}(r) \equiv \{A(r),\,B(r),\,X(r),\,\psi'(r),\ldots\}
\end{equation*}
collect the background fields entering the static, spherically symmetric ansatz. The exact black-hole background equations derived from the lifted parent action may be written schematically as
\begin{equation}
\bm{\mathcal{E}}\bigl[\bm{Y}(r),\bm{f}\bigr] = 0,
\label{eq:bg_equations}
\end{equation}
subject to horizon regularity and the asymptotic matching conditions in the overlap shell \(r_g\ll r\ll H_*^{-1}\) \cite{MukohyamaYingcharoenrat2022}.

Around a fiducial background \(\vev{\bm{Y}}(r)\) we linearize Eq.~\eqref{eq:bg_equations} as a first-order radial system,
\begin{equation*}
\frac{\dd}{\dd r}\,\delta \bm{Y}(r)
= \mathbf{M}(r)\,\delta \bm{Y}(r) + \mathbf{N}(r)\,\delta \bm{f},
\end{equation*}
where \(\mathbf{M}\) and \(\mathbf{N}\) are determined by the background equations and the chosen covariant basis. Introducing the fundamental transport matrix \(\mathbf{T}(r,r_m)\), with \(r_m\) a matching radius in the overlap region, the formal solution is
\begin{equation}
\delta \bm{Y}(r)
= \mathbf{T}(r,r_m)\,\delta \bm{Y}(r_m)
+ \int_{r_m}^{r} \dd r'\,\mathbf{T}(r,r')\,\mathbf{N}(r')\,\delta \bm{f}.
\label{eq:Ysolution}
\end{equation}
The first term propagates uncertainty in the asymptotic matching data, while the second transports genuine theory uncertainty encoded in the lifted coefficients. Since the ringdown timescale is much shorter than the Hubble time, the matching data at \(r_m\) can be treated as frozen at \(t_*\), with any residual cosmological drift suppressed by powers of \(r_g H_*\ll 1\). In the benchmark implementation we choose \(r_m\) inside the overlap shell and assess robustness by varying it over an order-unity range in \(r_m/r_g\); the diagnostics reported in Sec.~\ref{sec:discussion} should be interpreted as the stress-test envelope of that choice rather than as the output of a unique preferred matching radius.

The Einstein-frame arbitrary-background EFT coefficients are then functionals of both \(\bm{Y}(r)\) and \(\bm{f}\),
\begin{equation*}
\mathcal{C}_I^{E}(r)
= \mathcal{T}_I\circ\mathcal{D}_I\bigl[\bm{Y}(r),\bm{f}\bigr].
\end{equation*}
Linearizing about the fiducial background yields
\begin{equation}
\delta \mathcal{C}_I^{E}(r)
= \sum_{\alpha}\mathcal{S}_{I\alpha}(r)\,\delta f_{\alpha}
+ \sum_{p}\mathcal{B}_{Ip}(r)\,\delta Y_p(r),
\label{eq:Clinear1}
\end{equation}
where
\begin{equation*}
\mathcal{S}_{I\alpha}(r)
\equiv \left.\frac{\partial \mathcal{C}_I^{E}}{\partial f_{\alpha}}\right|_{\rm fid},
\qquad
\mathcal{B}_{Ip}(r)
\equiv \left.\frac{\partial \mathcal{C}_I^{E}}{\partial Y_p}\right|_{\rm fid}.
\end{equation*}
Substituting Eq.~\eqref{eq:Ysolution} into Eq.~\eqref{eq:Clinear1} we obtain a purely lifted-theory response,
\begin{equation}
\delta \mathcal{C}_I^{E}(r)
= \sum_{\alpha}\widetilde{\mathcal{S}}_{I\alpha}(r)\,\delta f_{\alpha}
+ \sum_{p}\widetilde{\mathcal{B}}_{Ip}(r)\,\delta Y_p(r_m),
\label{eq:Clinear2}
\end{equation}
with kernels \(\widetilde{\mathcal{S}}\) and \(\widetilde{\mathcal{B}}\) that encode the full radial transport.

\subsection{Inherited, anisotropy-activated, and strong-field sectors}
\label{subsec:classification}

Equation~\eqref{eq:Clinear2} makes it natural to decompose the black-hole EFT coefficients into three contributions,
\begin{equation*}
\delta \mathcal{C}_I^{E}(r)
= \delta \mathcal{C}_{I,\rm inh}^{E}(r)
+ \delta \mathcal{C}_{I,\rm aniso}^{E}(r)
+ \delta \mathcal{C}_{I,\rm sf}^{E}(r).
\end{equation*}
The first term is inherited directly from the isotropic FLRW projection and would remain even if all traceless background tensors vanished. The second is induced by the same parent jet but multiplies tensors such as \(\vev\sigma_{\mu\nu}\) and \(\vev r_{\mu\nu}\), so it is invisible on FLRW yet nonzero on a black-hole background. The third consists of genuinely strong-field coefficients not determined by the truncated lift and therefore to be constrained directly by black-hole observations.

The operator proportional to \(M_6\) is the prototype of the second class. On FLRW, where \(\vev\sigma^{\mu}{}_{\nu}=0\), it does not contribute to cosmological gravitational-wave propagation. On a static, spherically symmetric black-hole background, however, it modifies the effective kinetic structure of odd-parity perturbations and can split the radial and angular propagation speeds in the vicinity of the hole \cite{MukohyamaSerailleTakahashiYingcharoenrat2025}. This observation underpins the analysis: cosmological luminality does not, by itself, imply GR-like ringdown.

\subsection{Projection onto parity-resolved perturbation operators}
\label{subsec:projection_parity}

The transported coefficient set is now projected onto the parity sectors. We define the odd- and even-sector control vectors by
\begin{align*}
\delta \bm{\Theta}_{\rm odd}(r)
&= \mathbf{P}_{\rm odd}(r)\,\delta \bm{\mathcal{C}}^{E}(r),
\nonumber\\
\delta \bm{\Theta}_{\rm even}(r)
&= \mathbf{P}_{\rm even}(r)\,\delta \bm{\mathcal{C}}^{E}(r),
\end{align*}
where \(\mathbf{P}_{\rm odd}\) and \(\mathbf{P}_{\rm even}\) are algebraic projectors determined by the harmonic decomposition of the perturbations. In the odd sector one may choose
\begin{equation*}
\bm{\Theta}_{\rm odd}(r)
= \{Z_{\rm odd}(r),\,c_{r,\rm odd}^2(r),\,c_{\Omega,\rm odd}^2(r),\,V_{\rm odd}(r)\},
\end{equation*}
so that the generalized Regge-Wheeler operator takes the form
\begin{equation*}
\hat{\mathcal{L}}_{\ell}^{\rm odd}
= \frac{\dd^2}{\dd r_*^2} + \omega^2 - V_{\rm odd}\bigl[r;\bm{\Theta}_{\rm odd}(r)\bigr].
\end{equation*}
At the linearized level,
\begin{equation*}
\delta V_{\rm odd}(r) = \sum_I \mathcal{P}^{\rm odd}_{I}(r)\,\delta \mathcal{C}_I^{E}(r),
\end{equation*}
and similarly for the kinetic and propagation-speed coefficients. In particular, the anisotropic response of the odd tensor mode can be written schematically as
\begin{equation*}
\Delta c_{T,\rm odd}^2(r)
\equiv c_{r,\rm odd}^2(r)-c_{\Omega,\rm odd}^2(r)
\propto M_6(r)\,\vev\sigma(r),
\end{equation*}
which makes explicit the near-zone nature of the \(M_6\) effect.

In the even sector the perturbations obey a coupled operator equation,
\begin{equation*}
\hat{\mathcal{L}}_{\ell}^{\rm even}\bigl[\bm{\Theta}_{\rm even}(r)\bigr]\,\bm{u}_{\ell m}(r) = 0,
\end{equation*}
where \(\bm{u}_{\ell m}\) contains the generalized Zerilli field and the scalar fluctuation. In the minimal scordatura-compatible subclass studied recently, the scalar mode decouples from the metric perturbations at leading order in the scordatura parameter for monopole perturbations around an approximately stealth Schwarzschild solution \cite{MukohyamaTakahashiTomikawaYingcharoenrat2025}. This motivates a perturbative organization,
\begin{equation*}
\hat{\mathcal{L}}_{\ell}^{\rm even}
= \hat{\mathcal{L}}_{\ell,0}^{\rm even} + \alpha_L\,\delta \hat{\mathcal{L}}_{\ell,1}^{\rm even} + \cdots,
\end{equation*}
with \(\alpha_L\) the scordatura bookkeeping parameter. For our purposes the important point is that both parity sectors can be written as linear functionals of the transported coefficient vector \(\delta \bm{\mathcal{C}}^{E}(r)\).

\subsection{Response kernel for quasinormal modes}
\label{subsec:qnm_kernel}

The transported EFT coefficients are next propagated to QNM frequencies. Consider first the odd sector and let
\begin{equation*}
\hat{\mathcal{L}}_{\ell}^{\rm odd}(\omega)\,\Psi_{\ell n}^{R} = 0
\end{equation*}
be the right QNM eigenproblem with outgoing boundary conditions, and let \(\Psi_{\ell n}^{L}\) denote the corresponding adjoint mode. The first-order shift of the complex QNM frequency is then
\begin{equation}
\delta \omega_{\ell n}
= -\frac{\left\langle\!\left\langle \Psi_{\ell n}^{L},\,\delta \hat{\mathcal{L}}_{\ell}^{\rm odd}\,\Psi_{\ell n}^{R}\right\rangle\!\right\rangle}
{\left\langle\!\left\langle \Psi_{\ell n}^{L},\,(\partial_{\omega}\hat{\mathcal{L}}_{\ell}^{\rm odd})\,\Psi_{\ell n}^{R}\right\rangle\!\right\rangle}\Bigg|_{\omega=\omega_{\ell n}^{(0)}},
\label{eq:qnm_shift_formula}
\end{equation}
where \(\langle\!\langle \cdot,\cdot \rangle\!\rangle\) is the appropriate bilinear product for the non-self-adjoint QNM problem. Since \(\delta \hat{\mathcal{L}}_{\ell}^{\rm odd}\) is linear in \(\delta \mathcal{C}_I^{E}(r)\), Eq.~\eqref{eq:qnm_shift_formula} can be written as a radial response integral,
\begin{equation}
\delta \omega_{\ell n}
= \sum_I \int \dd r\,\mathcal{K}_{\ell n}^{I}(r)\,\delta \mathcal{C}_I^{E}(r),
\label{eq:qnm_kernel_radial}
\end{equation}
with \(\mathcal{K}_{\ell n}^{I}(r)\) a QNM kernel determined by the fiducial background and fiducial eigenfunctions.

Substituting Eq.~\eqref{eq:Clinear2} into Eq.~\eqref{eq:qnm_kernel_radial} gives the response to the lifted covariant coefficients,
\begin{equation}
\delta \omega_{\ell n}
= \sum_{\alpha}\Xi_{\ell n,\alpha}\,\delta f_{\alpha}
+ \sum_p \Xi^{(m)}_{\ell n,p}\,\delta Y_p(r_m),
\label{eq:qnm_f_response}
\end{equation}
where
\begin{equation*}
\Xi_{\ell n,\alpha}
\equiv \sum_I\int \dd r\,\mathcal{K}_{\ell n}^{I}(r)\,\widetilde{\mathcal{S}}_{I\alpha}(r).
\end{equation*}
Combining Eq.~\eqref{eq:qnm_f_response} with the lift matrix in Eq.~\eqref{eq:Lmatrix} yields the desired cosmology-to-ringdown response,
\begin{align}
\delta \omega_{\ell n}
&= \sum_i \Upsilon_{\ell n,i}\,\delta y_i
+ \sum_p \Xi^{(m)}_{\ell n,p}\,\delta Y_p(r_m),
\nonumber\\
\Upsilon_{\ell n,i}
&\equiv \sum_{\alpha}\Xi_{\ell n,\alpha}L_{\alpha i}.
\label{eq:Upsilon_def}
\end{align}
Equation~\eqref{eq:Upsilon_def} is the central practical result of the section. It shows that, within the chosen truncation, cosmological posterior samples can be pushed forward directly to QNM deviations by matrix multiplication and radial quadrature. The same construction applies to damping times, frequency ratios, or parity-resolved combinations of modes.

For the even sector the structure is analogous, except that the QNM problem is matrix valued and the kernel \(\mathcal{K}_{\ell n}^{I}(r)\) becomes a matrix of adjoint-mode overlaps. Since the even sector is presently under active development, we keep the operator notation abstract. Nevertheless, Eq.~\eqref{eq:Upsilon_def} remains valid after promoting the scalar kernel to its matrix generalization.

\subsection{Propagating cosmological constraints to spectroscopy space}
\label{subsec:constraints_to_spec}

Once the response matrix \(\Upsilon\) is known, cosmological covariance propagates to the black-hole observables according to
\begin{equation*}
\mathbf{C}_{\omega} = \Upsilon\,\mathbf{C}_{y}\,\Upsilon^{\mathsf T},
\end{equation*}
up to the contribution from matching uncertainties encoded in the second term of Eq.~\eqref{eq:Upsilon_def}. More generally, if \(\bm{q}_{\rm RD}\) denotes a chosen vector of ringdown observables, such as fractional shifts of the real and imaginary parts of QNM frequencies, then
\begin{equation*}
\delta \bm{q}_{\rm RD} = \mathbf{W}\,\delta \bm{y}_*,
\end{equation*}
for some response matrix \(\mathbf{W}\) assembled from the mode-by-mode kernels.

This leads to a clean definition of the allowed spectroscopy region,
\begin{equation}
\mathcal{A}_{\rm RD}
\equiv \Bigl\{\delta \bm{q}_{\rm RD}\;\big|\;\bm{y}_*\in\mathcal{A}_{\rm cos},\;\mathcal{S}_{\rm phys}[\bm{f},\bm{Y}] = 1\Bigr\},
\label{eq:Ard}
\end{equation}
where \(\mathcal{A}_{\rm cos}\) is the cosmologically allowed set and \(\mathcal{S}_{\rm phys}\) is a selection function implementing the non-cosmological consistency conditions: existence of a regular black-hole background, absence of ghosts and gradient instabilities in the relevant parity sector, positivity of the kinetic coefficients, and compatibility of the overlap matching. The corresponding no-go region is the complement of \(\mathcal{A}_{\rm RD}\) in the chosen spectroscopy plane.

Equation~\eqref{eq:Ard} is the key geometric point. The cosmological tensor-speed bound removes the homogeneous FLRW deformation but not anisotropy-activated operators such as the \(M_6\) sector. These operators are therefore natural targets for black-hole spectroscopy in cosmologically viable modified gravity \cite{MukohyamaSerailleTakahashiYingcharoenrat2025}: ringdown probes the inhomogeneous completion of the same EFT that cosmology constrains in its isotropic limit.

\section{Phenomenological forecasts and allowed/no-go regions for ringdown observables}
\label{sec:phenom}

The formal construction of Secs.~\ref{sec:framework} and~\ref{sec:map} becomes phenomenological only after it is paired with a detector model. The forecast concerns not arbitrary ringdown deformations, but the subset that survives cosmological viability and transport to the black-hole near zone. This gives a spectroscopy forecast conditioned by cosmology rather than a theory-agnostic deformation analysis \cite{Dreyer2004,BertiCardosoWill2006}.

The forecast uses four ingredients: a ringdown observable basis, a detector likelihood, the cosmology-conditioned prior induced by the EFT transport, and the corresponding allowed, excluded, and undetectable regions in spectroscopy space. The same setup can be applied to current detectors, third-generation observatories such as Einstein Telescope, and space-based missions such as LISA \cite{MaggioreEtAl2020,AmaroSeoaneEtAl2017}.

\subsection{Forecast basis and detector likelihood}
\label{subsec:forecast_basis}

A convenient ringdown forecast basis is the vector of fractional shifts of the real and imaginary parts of the dominant QNM frequencies,
\begin{equation}
\bm{q}
\equiv
\begin{pmatrix}
\delta \omega^{\rm R}_{220}/\omega^{\rm R}_{220,{\rm GR}} \\
\delta \omega^{\rm I}_{220}/\omega^{\rm I}_{220,{\rm GR}} \\
\delta \omega^{\rm R}_{330}/\omega^{\rm R}_{330,{\rm GR}} \\
\delta \omega^{\rm I}_{330}/\omega^{\rm I}_{330,{\rm GR}} \\
\vdots
\end{pmatrix},
\label{eq:qbasis}
\end{equation}
where the ellipsis indicates whichever subset of overtones or subdominant angular multipoles is measurable for the detector and source population under consideration. In a parity-resolved forecast one may equivalently choose a basis adapted to odd and even sectors, but Eq.~\eqref{eq:qbasis} is sufficient for the general discussion.

The map derived in Sec.~\ref{sec:map} implies that the ringdown variables are not free parameters but linear functionals of the cosmological data and of residual nuisance directions associated with the matching shell or unresolved strong-field coefficients. We therefore write
\begin{equation}
\delta \bm{q}
=
\mathbf{W}\,\delta \bm{y}_*
+
\mathbf{W}_{\eta}\,\delta \bm{\eta},
\qquad
\delta \bm{\eta}
\equiv
\{\delta \bm{Y}(r_m),\,\delta \bm{c}_{\rm sf}\},
\label{eq:qmap_nuisance}
\end{equation}
where \(\delta \bm{y}_*\) is the cosmological input vector introduced in Eq.~\eqref{eq:yvec}, \(\delta \bm{Y}(r_m)\) denotes uncertainty in the matching data at the overlap radius, and \(\delta \bm{c}_{\rm sf}\) collects genuinely strong-field coefficients retained as nuisance directions. Equation~\eqref{eq:qmap_nuisance} is the phenomenological compression of the full EFT transport.

Within the detector-metric construction used to interpret the Bayesian products, we start from the standard network inner product
\begin{equation*}
(a|b)
\equiv
4\,\mathrm{Re}\sum_{X}
\int_{0}^{\infty}
\frac{\tilde a_X^{*}(f)\,\tilde b_X(f)}{S_{n,X}(f)}\,\dd f,
\end{equation*}
where \(X\) labels detectors and \(S_{n,X}(f)\) is the one-sided noise power spectral density. The ringdown signal-to-noise ratio is then
\begin{equation*}
\rho_{\rm RD}^{2} = (h_{\rm RD}|h_{\rm RD}).
\end{equation*}
For a local Gaussian summary of a given likelihood, if \(\bm{\vartheta}\) denotes the waveform parameters, including the remnant mass and spin, mode amplitudes, start time, phase, and the spectroscopy vector \(\bm{q}\), the curvature matrix is
\begin{equation*}
\Gamma_{ab}
=
\left(\frac{\partial h_{\rm RD}}{\partial \vartheta_a}\Bigg|\frac{\partial h_{\rm RD}}{\partial \vartheta_b}\right).
\end{equation*}
After marginalizing over all nuisance waveform parameters except \(\bm{q}\), one obtains the detector covariance
\begin{equation*}
\mathbf{C}_{\rm det}
=
\bigl(\mathbf{\Gamma}_{qq}^{\rm marg}\bigr)^{-1}.
\end{equation*}
For fixed detector noise curve and waveform family, \(\mathbf{C}_{\rm det}\propto \rho_{\rm RD}^{-2}\) in the high-SNR Gaussian limit. In the implementation below, this covariance is not inserted by hand; it is read from the Bayesian injection products after marginalizing over linear mode amplitudes and then augmented, where indicated, by a separate remnant-calibration covariance.

\subsection{Cosmology-conditioned prior and combined posterior}
\label{subsec:prior_posterior}

The response map of Sec.~\ref{sec:map} immediately induces a prior covariance in spectroscopy space,
\begin{equation}
\mathbf{C}_{\rm prior}
=
\mathbf{W}\,\mathbf{C}_{y}\,\mathbf{W}^{\mathsf T}
+
\mathbf{W}_{\eta}\,\mathbf{C}_{\eta}\,\mathbf{W}_{\eta}^{\mathsf T},
\label{eq:Cprior}
\end{equation}
where \(\mathbf{C}_{y}\) is the covariance of the cosmological input vector and \(\mathbf{C}_{\eta}\) encodes the prior width assigned to matching and strong-field nuisance directions. Equation~\eqref{eq:Cprior} is the quantity that distinguishes the present framework from generic parameterized-ringdown analyses: the prior ellipsoid is not arbitrary but inherited from the cosmological posterior and filtered through the black-hole EFT map.

In the implemented compressed posterior, \(\mathbf C_y\) is estimated directly from the accepted samples in \texttt{data/posterior\_samples.csv}. The inherited component is evaluated through \(q_{\rm inh}=3\alpha_{T0}/2\), while the anisotropy component is drawn on the demonstrated positive Hayward branch and pushed forward with the response vector calibrated in Eq.~\eqref{eq:response_eta}. This sample-based pushforward produces the two derived rows in Table~\ref{tab:posterior_summary} and the two panels of Fig.~\ref{fig:cosmo_posterior_pushforward}. Thus the Gaussian expressions below are not merely formal: their covariance input is generated by an executable likelihood layer, while non-Gaussian checks can be made by using the same posterior samples directly.

Let \(\hat{\bm{q}}\) denote the maximum-likelihood ringdown estimate. Assuming Gaussian errors and expanding around a fiducial GR point, the combined posterior is
\begin{align*}
\mathbf{C}_{\rm post}
&=
\left(\mathbf{C}_{\rm prior}^{-1}+\mathbf{C}_{\rm det}^{-1}\right)^{-1},\\
\bar{\bm{q}}_{\rm post}
&=
\mathbf{C}_{\rm post}\,\mathbf{C}_{\rm det}^{-1}\hat{\bm{q}},
\end{align*}
where a nonzero prior mean may be restored straightforwardly if the fiducial theory point is not exactly GR\@. A useful tension statistic for the cosmology-conditioned EFT class is
\begin{equation}
\Delta\chi_{\rm EFT}^{2}
=
\hat{\bm{q}}^{\mathsf T}
\left(\mathbf{C}_{\rm det}+\mathbf{C}_{\rm prior}\right)^{-1}
\hat{\bm{q}}.
\label{eq:chiEFT}
\end{equation}
If \(\Delta\chi_{\rm EFT}^{2}\) exceeds the appropriate confidence threshold, the event is in tension with the cosmologically viable EFT family under consideration, rather than merely with GR in isolation.

Equations~\eqref{eq:Cprior}--\eqref{eq:chiEFT} also give the event-level likelihood layer for hierarchical inference: cosmology supplies a theory prior on ringdown deformations, and gravitational-wave data update it event by event. If the map from \(\bm{y}_*\) to \(\bm{q}\) is significantly nonlinear, direct posterior pushforward replaces the Gaussian expressions. The ellipsoidal description used here is the local approximation for analytic forecasts.

\subsection{Consistency eigenmodes and detectability thresholds}
\label{subsec:eigenmodes}

The comparison between cosmological viability and detector reach is most transparent in a detector-whitened basis. Define
\begin{equation*}
\mathbf{M}
\equiv
\mathbf{C}_{\rm det}^{-1/2}
\mathbf{C}_{\rm prior}
\mathbf{C}_{\rm det}^{-1/2},
\end{equation*}
and solve the eigenvalue problem
\begin{equation*}
\mathbf{M}\,\bm{u}_k = \lambda_k\,\bm{u}_k,
\qquad
\bm{u}_k^{\mathsf T}\bm{u}_j = \delta_{kj}.
\end{equation*}
Equivalently, in the original coordinates one may solve the generalized eigenvalue problem
\begin{equation*}
\mathbf{C}_{\rm prior}\,\bm{v}_k
=
\lambda_k\,\mathbf{C}_{\rm det}\,\bm{v}_k,
\qquad
\bm{v}_k \equiv \mathbf{C}_{\rm det}^{1/2}\bm{u}_k.
\end{equation*}
The eigenvectors \(\bm{v}_k\) define the principal directions in spectroscopy space selected jointly by the cosmological map and by the detector response. They are the natural variables in which to report the phenomenological reach of the theory class.

To interpret the eigenvalues, consider a displacement \(\delta \bm{q}=\epsilon\,\bm{v}_k\). In the whitened basis the detector significance is
\begin{equation*}
\chi_{\rm det}^{2}=\epsilon^{2},
\end{equation*}
while the cosmology-conditioned prior weight is
\begin{equation*}
\chi_{\rm prior}^{2}=\frac{\epsilon^{2}}{\lambda_k}.
\end{equation*}
At confidence level \(p\), corresponding to threshold \(\Delta\chi_{p}^{2}\), the maximum cosmology-allowed excursion and the minimum detector-resolvable excursion along mode \(k\) are therefore
\begin{equation*}
\epsilon^{\rm max}_{{\rm allow},k}(p)
=
\sqrt{\lambda_k\,\Delta\chi_{p}^{2}},
\qquad
\epsilon^{\rm min}_{{\rm det},k}(p)
=
\sqrt{\Delta\chi_{p}^{2}}.
\end{equation*}
This yields an immediate criterion,
\begin{equation}
\lambda_k > 1.
\label{eq:lambda_criterion}
\end{equation}
Equation~\eqref{eq:lambda_criterion} is equivalent to the statement that there exists a cosmologically allowed deviation along mode \(k\) that is detectable by the chosen ringdown measurement. Conversely, directions with \(\lambda_k<1\) are phenomenological no-go directions for that detector: the cosmology-conditioned prior confines them below the detector threshold.

The same construction gives a useful SNR threshold. If the detector covariance scales as
\begin{equation*}
\mathbf{C}_{\rm det}(\rho_{\rm RD}) = \rho_{\rm RD}^{-2}\,\bar{\mathbf{C}}_{\rm det},
\end{equation*}
then
\begin{equation*}
\lambda_k(\rho_{\rm RD}) = \rho_{\rm RD}^{2}\,\bar\lambda_k,
\end{equation*}
where \(\bar\lambda_k\) is computed from \(\bar{\mathbf{C}}_{\rm det}\). The critical ringdown SNR required to access mode \(k\) is therefore
\begin{equation}
\rho_{{\rm crit},k} = \bar\lambda_k^{-1/2}.
\label{eq:rho_crit}
\end{equation}
Equation~\eqref{eq:rho_crit} provides a compact detector-agnostic summary of the forecast: it tells us how loud the ringdown must be for a given cosmologically viable deformation direction to become observable.

\subsection{Allowed, excluded, and undetectable regions}
\label{subsec:allowed_regions}

The eigenmode analysis has a simple geometric interpretation. At confidence level \(p\), the cosmology-conditioned allowed region in spectroscopy space is
\begin{equation*}
\mathcal{A}_{p}
\equiv
\left\{
\delta \bm{q}
\;\middle|\;
\delta \bm{q}^{\mathsf T}\mathbf{C}_{\rm prior}^{-1}\delta \bm{q}
\le \Delta\chi_{p}^{2},
\ \mathcal{S}_{\rm phys}=1
\right\},
\end{equation*}
where \(\mathcal{S}_{\rm phys}\) implements the non-cosmological consistency conditions already discussed in Sec.~\ref{subsec:constraints_to_spec}: existence of a regular black-hole background, positivity of kinetic coefficients, hyperbolicity of the perturbation equations, and admissible overlap matching. The corresponding theory no-go region is simply
\begin{equation*}
\mathcal{N}_{p} \equiv \mathbb{R}^{n_q}\setminus \mathcal{A}_{p}.
\end{equation*}

For a detector or detector network \(X\), the \(p\)-level resolvable region is
\begin{equation*}
\mathcal{D}^{(X)}_{p}
\equiv
\left\{
\delta \bm{q}
\;\middle|\;
\delta \bm{q}^{\mathsf T}\bigl(\mathbf{C}^{(X)}_{\rm det}\bigr)^{-1}\delta \bm{q}
\ge \Delta\chi_{p}^{2}
\right\},
\end{equation*}
namely the exterior of the detector error ellipsoid centered on the fiducial point. This leads to a natural three-way partition of the spectroscopy plane:
\begin{align*}
\mathcal{V}^{(X)}_{p}
&\equiv \mathcal{A}_{p}\cap \mathcal{D}^{(X)}_{p},
&&\text{viable and detectable},\\
\mathcal{U}^{(X)}_{p}
&\equiv \mathcal{A}_{p}\setminus \mathcal{D}^{(X)}_{p},
&&\text{viable but undetectable},\\
\mathcal{N}_{p}
&= \mathbb{R}^{n_q}\setminus \mathcal{A}_{p},
&&\text{excluded / no-go}.
\end{align*}
In two-dimensional Gaussian projections these sets appear as ellipses and their complements. In a nonlinear treatment they are obtained by pushing posterior samples through the EFT map and constructing the corresponding highest-posterior-density regions.

A particularly useful visualization is obtained by projecting onto the plane
\begin{equation*}
\left(
\frac{\delta \omega^{\rm R}_{220}}{\omega^{\rm R}_{220,{\rm GR}}},
\frac{\delta \omega^{\rm I}_{220}}{\omega^{\rm I}_{220,{\rm GR}}}
\right),
\end{equation*}
for current detectors, and then promoting the basis to include \((330)\) or other subdominant modes for third-generation or space-based missions. When parity-resolved information can be tracked theoretically, an additional projection onto anisotropy-sensitive combinations isolates the contribution of operators such as the \(M_6\) sector.

\subsection{Bayesian multi-mode injection realization}
\label{subsec:bayesian_injections}

The detector side is evaluated with a reproducible Bayesian time-domain injection suite in \ringdowncode. The injected signal is a whitened damped-sinusoid model,
\begin{equation}
 h(t)=\sum_{a\in\{220,221,330\}}
 A_a e^{-\gamma_a(q_I)t}
 \cos\!\left[\omega_a(q_R)t+\phi_a\right] + n(t),
\label{eq:bayes_injection_waveform}
\end{equation}
where the seeded Gaussian noise realization is fixed, the representative remnant scale is held fixed in geometric units, and the injected deformation is the admissible Hayward point
\begin{equation*}
(q_R,q_I)_{\rm inj}=(-5.3\times10^{-3},\,1.5\times10^{-3}).
\end{equation*}
The parameters recovered by the likelihood are the fractional shift of the dominant-mode real frequency, \(q_R\), and the fractional shift of the damping rate, \(q_I\), in the same coordinates used by the EFT response vector of Eq.~\eqref{eq:response_eta}. For each grid point in \((q_R,q_I)\), the sine and cosine amplitudes of every included mode are analytically marginalized with broad Gaussian priors. This gives a normalized marginal posterior
\begin{equation*}
 p(q_R,q_I|d,t_{\rm start},{\cal M})
 \propto
 \int d\bm a\,
 \exp\!\left[-\frac12(d-h(q_R,q_I,\bm a)|d-h(q_R,q_I,\bm a))\right]p(\bm a),
\end{equation*}
for each mode set \({\cal M}\) and each ringdown-start-time cut.

\begin{figure}[t]
\centering
\includegraphics[width=0.94\textwidth]{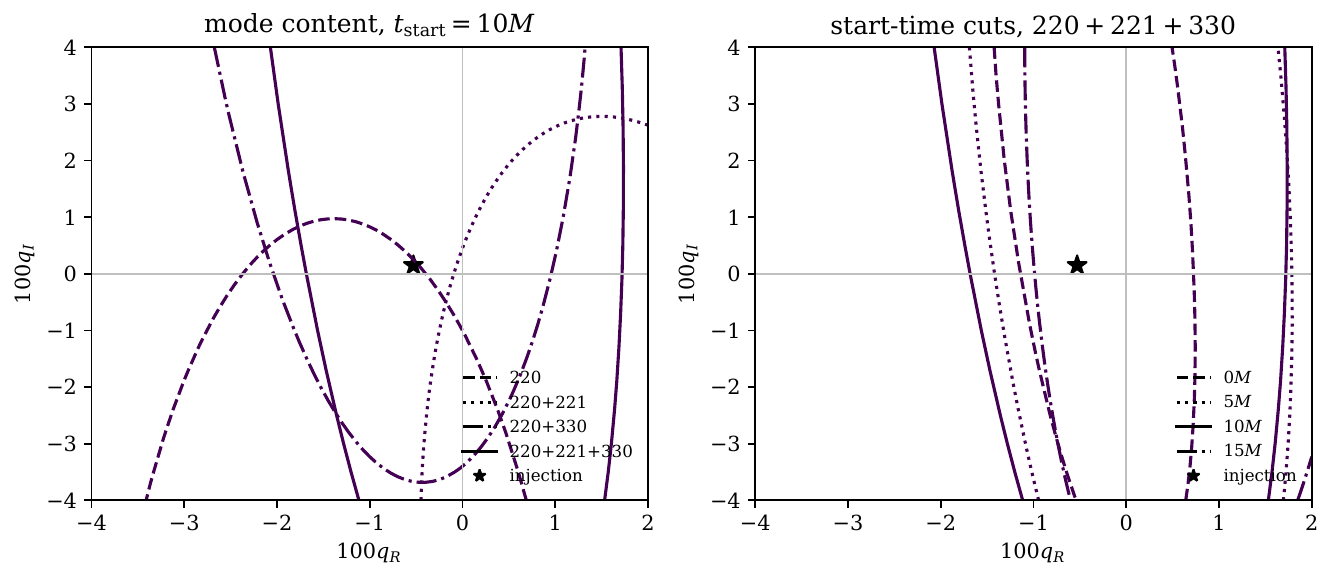}
\caption{Bayesian ringdown-injection posteriors for the detector-side part of the map. The left panel compares one-, two-, and three-mode recovery models at fixed \(t_{\rm start}=10M\); the right panel repeats the full three-mode recovery after cutting the signal at \(t_{\rm start}/M=0,5,10,15\). Contours enclose 90\% posterior probability in the \((q_R,q_I)\) plane. The injected deformation is the admissible Hayward point, and linear sine/cosine amplitudes are analytically marginalized at each grid point.}
\label{fig:bayes_ringdown_posteriors}
\end{figure}

\begin{figure}[t]
\centering
\includegraphics[width=0.68\textwidth]{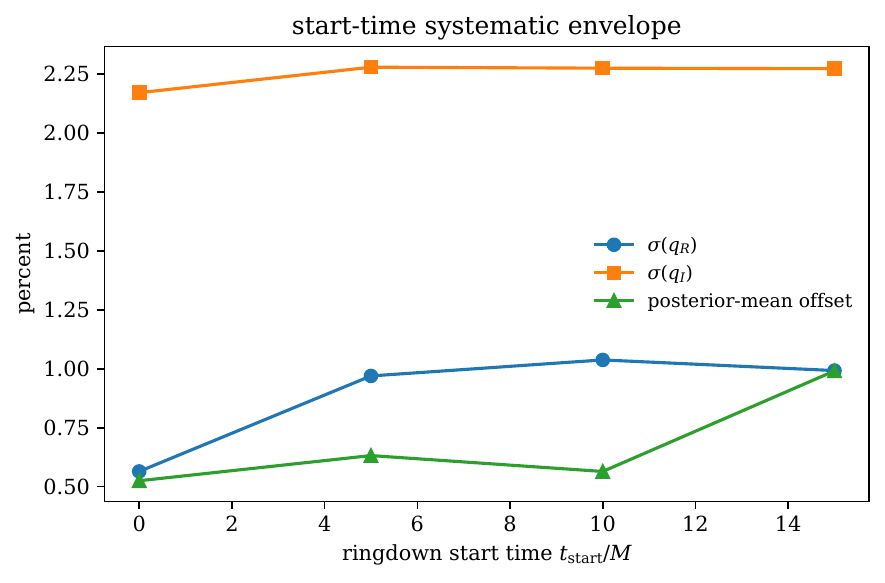}
\caption{Ringdown-start-time systematic envelope for the full three-mode Bayesian recovery. The plot shows the one-sigma posterior widths in \(q_R\) and \(q_I\), together with the posterior-mean displacement from the injected value, as the start time is varied. The envelope is used as a detector-side systematic check when interpreting anisotropy-activated signals.}
\label{fig:start_time_systematics}
\end{figure}

The executable analysis explicitly tests two detector-side systematics that a single covariance cannot test. First, it compares one-, two-, and three-mode recoveries: \((220)\), \((220+221)\), \((220+330)\), and \((220+221+330)\). Second, it repeats the recovery for \(t_{\rm start}/M\in\{0,5,10,15\}\). The remnant-calibration effect is retained separately in the covariance ladder used for the consistency-eigenmode calculation; the injection test is instead designed to isolate mode truncation and start-time uncertainty. Table~\ref{tab:ringdown_bayes} and Figs.~\ref{fig:bayes_ringdown_posteriors}--\ref{fig:start_time_systematics} summarize the resulting posterior widths, correlations, and systematic shifts. The key practical lesson is that the admissible anisotropy signal sits at the level where early-time mode truncation can matter: the full multi-mode recovery is more stable under changes of \(t_{\rm start}\), while truncated recoveries can move the posterior median by an amount comparable to the statistical width.

\begin{table}[H]
\caption{Bayesian ringdown injection/recovery summary generated by \ringdowncode. The injected value is \((q_R,q_I)=(-0.53\%,+0.15\%)\). The first four rows compare recovery mode content at fixed \(t_{\rm start}=10M\); the last three rows show the start-time systematic for the full three-mode recovery. \(A_{90}\) is the area of the two-dimensional highest-posterior-density region enclosing 90\% probability.}
\label{tab:ringdown_bayes}
\centering
\small
\begin{tabular*}{\textwidth}{@{\extracolsep{\fill}}l c c c c c c@{}}
\hline\hline
Modes & $t_{\rm start}/M$ & $\rho_{\rm RD}$ & $q_R$ [\%] & $q_I$ [\%] & Corr. & $A_{90}$ \\
\hline
220 & 10 & 100 & $-1.36\pm0.86$ & $-2.40\pm1.48$ & $-0.02$ & $1.44e-03$ \\
220+221 & 10 & 100 & $+1.07\pm0.73$ & $-1.51\pm1.90$ & $-0.10$ & $1.42e-03$ \\
220+330 & 10 & 100 & $-0.59\pm0.92$ & $+1.42\pm1.99$ & $-0.07$ & $2.20e-03$ \\
220+221+330 & 10 & 100 & $-0.10\pm1.04$ & $+0.51\pm2.27$ & $-0.09$ & $2.71e-03$ \\
220+221+330 & 0 & 100 & $-0.25\pm0.56$ & $+0.59\pm2.17$ & $-0.25$ & $1.43e-03$ \\
220+221+330 & 5 & 100 & $+0.07\pm0.97$ & $+0.35\pm2.28$ & $-0.10$ & $2.54e-03$ \\
220+221+330 & 15 & 100 & $+0.42\pm0.99$ & $+0.45\pm2.27$ & $+0.04$ & $2.41e-03$ \\
\hline\hline
\end{tabular*}

\end{table}

\subsection{Detector covariance ladder from Bayesian injections}
\label{subsec:bayes_cov_ladder}

The same injection script evaluates the three-mode posterior at three reference ringdown signal-to-noise ratios, denoted current-like, ET-like, and LISA-like. These labels are not mission forecasts; they are reproducible covariance products that define increasingly precise detector layers in the same \((q_R,q_I)\) coordinates as the EFT response map. At the fiducial start time \(t_{\rm start}=10M\), the response-direction one-sigma resolutions along \(\bm W_\eta\) are approximately \(2.47\%\), \(1.53\%\), and \(0.55\%\), respectively. The direct admissible point is displaced by \(0.55\%\) along this direction, while the demonstrated admissible endpoint is displaced by \(1.54\%\); the proxy continuation reaches larger displacements and is therefore a detector-reach benchmark rather than an established viability claim.

The companion file \texttt{data/ringdown\_bayesian\_summary.csv} stores the full posterior means, standard deviations, correlations, 90\% highest-posterior-density areas, and consistency-eigenvalue products used below. The geometry of Eqs.~\eqref{eq:Cprior}--\eqref{eq:lambda_criterion} is exact within the local Gaussian compression of this posterior, while Figs.~\ref{fig:bayes_ringdown_posteriors}--\ref{fig:start_time_systematics} keep the non-Gaussian contour and start-time information available as an explicit systematic check.

\section{Concrete worked example and literature-calibrated case studies}
\label{sec:benchmarks}

This section applies the map. The stealth-Schwarzschild branch provides a complete odd-parity example from cosmological luminality to a ringdown shift. Proposition~1 is a structural EFT statement. The Hayward calculations calibrate the size and direction of the anisotropy-activated odd sector, without being treated as a newly sampled strong-field posterior. The detector comparison uses the Bayesian multi-mode covariance and start-time envelope of Sec.~\ref{subsec:bayesian_injections}. The section closes with a weak-mixing even-sector result and a slow-spin continuation of the inherited branch. The strongest quantitative statements remain those for odd, nonspinning perturbations; results beyond that baseline are labeled as extensions or deferred work.

\subsection{Inherited stealth-Schwarzschild branch}
\label{subsec:worked_stealth}

Consider the shift- and reflection-symmetric odd-parity sector of quadratic higher-order scalar-tensor (DHOST-type) theories for which the arbitrary-background EFT coefficients satisfy
\begin{equation}
\alpha(r)+M_3^2(r)=0.
\label{eq:p4zero_condition}
\end{equation}
In the generalized Regge--Wheeler construction this condition sets $p_4=0$ and removes the problematic odd-sector mixing that otherwise spoils asymptotic regularity of slowly rotating solutions \cite{MukohyamaTakahashiYingcharoenrat2022,MukohyamaTakahashiTomikawaYingcharoenrat2023}. On the stealth-Schwarzschild branch,
\begin{align}
A(r)=B(r)&=1-\frac{r_H}{r},
\nonumber\\
M_*^2(r)&=M_*^2={\rm const},
\qquad
M_3^2(r)=M_3^2={\rm const}.
\label{eq:stealth_branch}
\end{align}
The covariant lift of Sec.~\ref{sec:map} therefore truncates already at zeroth order in the odd sector:
\begin{equation*}
J^0\mathcal F
\longrightarrow
\left\{
M_*^2,\,
M_3^2,\,
\alpha=-M_3^2
\right\}.
\end{equation*}
The transported black-hole EFT coefficients are therefore radially constant in the odd sector, and the two odd-mode sound speeds coincide,
\begin{align*}
c_\rho^2 = c_\theta^2 \equiv c_T^2
&= \frac{M_*^2}{M_*^2+M_3^2}
= 1+\alpha_T,
\\
\alpha_T
&= - \frac{M_3^2}{M_*^2+M_3^2},
\end{align*}
with the stability conditions requiring
\begin{equation*}
M_*^2>0,
\qquad
M_*^2+M_3^2>0
\quad\Longleftrightarrow\quad
1+\alpha_T>0.
\end{equation*}

For constant $\alpha_T$, the odd master equation can be mapped exactly to the GR Regge--Wheeler equation by the rescalings
\begin{equation*}
\tilde\omega = (1+\alpha_T)^{-1/2}\omega,
\qquad
\tilde r_* = \sqrt{1+\alpha_T}\,r_*,
\end{equation*}
which implies the closed-form QNM relation
\begin{equation}
r_H \omega_{\ell n}
=
r_H \omega_{\ell n}^{\rm GR}(1+\alpha_T)^{3/2}.
\label{eq:qnm_scaling_stealth}
\end{equation}
For the fundamental odd mode $(\ell,n)=(2,0)$, using
\begin{equation*}
r_H \omega_{220}^{\rm GR}=0.747343-0.177925\,\ii,
\end{equation*}
we obtain
\begin{equation*}
\delta(r_H \omega_{220})
=
\left(1.121014-0.266888\,\ii\right)\alpha_T
+\Order{\alpha_T^2},
\end{equation*}
or equivalently
\begin{equation}
\frac{\delta\omega^{\rm R}_{220}}{\omega^{\rm R}_{220,{\rm GR}}}
=
\frac{\delta|\omega^{\rm I}_{220}|}{|\omega^{\rm I}_{220,{\rm GR}}|}
=
(1+\alpha_T)^{3/2}-1
=
\frac{3}{2}\alpha_T+\Order{\alpha_T^2}.
\label{eq:relative_shift_stealth}
\end{equation}
Equation~\eqref{eq:qnm_scaling_stealth} is the first explicit end-to-end output of the map. The cosmological lift fixes the asymptotic odd-sector parameter $\alpha_T$, black-hole EFT transport reduces the odd master equation to a rescaling, and the pushforward to the $(220)$ QNM shift is analytic. The calculation then consists of identifying \(\alpha_T\) on the luminal stealth branch, applying Eq.~\eqref{eq:qnm_scaling_stealth}, and imposing the multimessenger bound in Eq.~\eqref{eq:gw170817_speed}.

The observational meaning of Eq.~\eqref{eq:qnm_scaling_stealth} is slightly subtler than the fixed-\(r_H\) notation suggests. Because every odd-parity mode rescales by the same factor, the entire inherited odd spectrum is equivalent to a GR Schwarzschild spectrum with an effective mass
\begin{equation*}
M_{\rm eff}=M(1+\alpha_T)^{-3/2},
\qquad
\omega_{\ell n}^{\rm inh}(M)=\omega_{\ell n}^{\rm GR}(M_{\rm eff}).
\end{equation*}
All odd-sector mode ratios therefore remain GR-like, so odd-parity ringdown alone cannot determine \(\alpha_T\) unless the remnant mass scale is supplied independently. The cosmological prior then freezes even this externally mass-calibrated inherited branch almost completely. The multimessenger observation of GW170817/GRB170817A constrains the speed difference between gravity and light to \cite{AbbottEtAl2017Multimessenger}
\begin{equation}
-3\times10^{-15}
<
\frac{c_T}{c}-1
<
7\times10^{-16},
\label{eq:gw170817_speed}
\end{equation}
which implies
\begin{equation}
|\alpha_T| = |c_T^2-1| \lesssim 6\times10^{-15}.
\label{eq:alphaT_bound_explicit}
\end{equation}
If the remnant mass is calibrated independently, substituting Eq.~\eqref{eq:alphaT_bound_explicit} into Eq.~\eqref{eq:relative_shift_stealth} yields
\begin{equation}
\left|
\frac{\delta\omega_{220}}{\omega_{220,{\rm GR}}}
\right|_{\rm inh}
\lesssim 9\times10^{-15}.
\label{eq:inherited_bound_final}
\end{equation}
The inherited odd-parity branch is therefore a double null result: odd-parity ringdown absorbs it into a mass redefinition, and cosmology suppresses any externally mass-calibrated residual far below foreseeable sensitivity. This applies only to the inherited isotropic odd branch tied to the asymptotic FLRW tensor-speed parameter; it does not bound anisotropy-activated coefficients such as \(M_6\).

\begin{table}[H]
\caption{Fully worked inherited odd-parity branch on the stealth-Schwarzschild background. The first four rows reproduce the $(\ell,n)=(2,0)$ values at fixed horizon radius, obtained from the exact scaling law~\eqref{eq:qnm_scaling_stealth} and from direct integration in Ref.~\cite{MukohyamaTakahashiTomikawaYingcharoenrat2023}. Because every odd-parity mode rescales by the same factor, the whole branch is spectroscopically degenerate with a mass redefinition; the last row therefore gives the cosmologically allowed interval for the corresponding rescaling once the remnant mass is calibrated independently through GW170817/GRB170817A.}
\label{tab:stealth220}
\begin{tabular}{c c c}
\(\alpha_T\) & \(r_H \omega_{220}\) & \(\delta \omega_{220}/\omega_{220,{\rm GR}}\) \\
\hline
0 & \(0.747343 - 0.177925\,\ii\) & 0 \\
0.04 & \(0.792629 - 0.188706\,\ii\) & \(+6.06\%\) \\
0.08 & \(0.838795 - 0.199697\,\ii\) & \(+12.24\%\) \\
0.12 & \(0.885824 - 0.210893\,\ii\) & \(+18.53\%\) \\
\(|\alpha_T| \lesssim 6\times10^{-15}\) & cosmologically allowed branch & \(\lesssim 9\times10^{-15}\) \\
\end{tabular}
\end{table}

\subsection{Proposition: anisotropy-activated luminality loophole}
\label{subsec:luminality_loophole}

\paragraph*{Proposition 1 (Existence of an anisotropy-activated luminality loophole).}
Consider the arbitrary-background timelike-scalar EFT on a static, spherically symmetric black-hole background with a timelike scalar profile, nonvanishing traceless background tensors in the near zone, and an asymptotically luminal FLRW sector. Then cosmological luminality on FLRW does not imply luminal odd-parity propagation on the black-hole background. In particular, operators multiplying traceless background tensors, exemplified by
\begin{equation}
M_6(r)\,\bar\sigma^\mu{}_{\nu}\,
\delta K^{\nu}{}_{\alpha}\delta K^{\alpha}{}_{\mu},
\label{eq:M6_operator_repeat}
\end{equation}
are invisible in the homogeneous cosmological sector but can generate finite $\alpha_T^{(\rho)}\neq \alpha_T^{(\theta)}$ and hence finite odd-sector QNM shifts in the black-hole near zone.

\emph{Proof.}
On FLRW, spatial isotropy implies
\begin{equation*}
\bar\sigma^\mu{}_{\nu} = 0,
\qquad
\bar r^\mu{}_{\nu} = 0,
\end{equation*}
so every operator proportional to a traceless background tensor vanishes identically in the cosmological sector. In particular, the $M_6$ operator in Eq.~\eqref{eq:M6_operator_repeat} does not contribute to the homogeneous tensor-speed observable constrained by $\Pi_T$. By contrast, on a static and spherically symmetric background with a timelike scalar profile, $\bar\sigma^\mu{}_{\nu}\neq0$ generically, and the odd master equation depends separately on the radial and angular propagation functions $\alpha_T^{(\rho)}$ and $\alpha_T^{(\theta)}$ \cite{MukohyamaSerailleTakahashiYingcharoenrat2025}.

A concrete asymptotically luminal example is provided by the Hayward branch studied in Ref.~\cite{MukohyamaSerailleTakahashiYingcharoenrat2025}. Imposing asymptotic luminality fixes the EFT branch so that, in terms of $x=r/r_g$ and $\eta_{\rm BH}\equiv \sigma^3/r_g^3$, one has
\begin{equation}
\begin{aligned}
\alpha_T^{(\rho)}(x)
&=
\frac{2\eta_{\rm BH}(2x^3+\eta_{\rm BH})}{x^6-4\eta_{\rm BH} x^3-2\eta_{\rm BH}^2},
\\
\alpha_T^{(\theta)}(x)
&=
-\frac{\eta_{\rm BH}(2x^3+\eta_{\rm BH})}{(x^3+\eta_{\rm BH})^2},
\end{aligned}
\label{eq:alphat_hayward}
\end{equation}
with
\begin{equation}
-1<\eta_{\rm BH}<\eta_c,
\qquad
\eta_c = \sqrt{\frac{3}{2}}-1 \simeq 0.224,
\label{eq:eta_range}
\end{equation}
and
\begin{equation*}
\begin{aligned}
M_6
&=
4M_*^2 r_g \eta_{\rm BH}
\sqrt{\frac{1-4\eta_{\rm BH}+2\eta_{\rm BH}^2}{1+\eta_{\rm BH}}}
\\
&\qquad\times
\frac{(x^3+\eta_{\rm BH})^{3/2}(2x^3+\eta_{\rm BH})}{x^9}.
\end{aligned}
\end{equation*}
By construction,
\begin{equation*}
\lim_{x\to\infty}\alpha_T^{(\rho)}(x)
=
\lim_{x\to\infty}\alpha_T^{(\theta)}(x)=0,
\end{equation*}
so the cosmological and asymptotic gravitational-wave-speed bound is respected. Nevertheless, at the black-hole horizon as it enters the odd-sector propagation coefficients, $x=1$,
\begin{equation}
\begin{aligned}
\alpha_T^{(\rho)}(1)
&=
\frac{2\eta_{\rm BH}(2+\eta_{\rm BH})}{1-4\eta_{\rm BH}-2\eta_{\rm BH}^2},
\\
\alpha_T^{(\theta)}(1)
&=
-\frac{\eta_{\rm BH}(2+\eta_{\rm BH})}{(1+\eta_{\rm BH})^2},
\end{aligned}
\label{eq:nearzone_split}
\end{equation}
which are nonzero for every $\eta_{\rm BH}\neq0$ in the stable interval~\eqref{eq:eta_range}. Therefore asymptotic cosmological luminality coexists with finite odd-sector nonluminality in the black-hole near zone. \hfill\(\square\)

\paragraph*{Corollary 1.}
The filtered cosmological prior projects onto a bounded inherited odd-sector branch and an unsuppressed anisotropy-activated branch:
\begin{equation*}
\delta \bm{q}_{220}^{\rm odd}
=
\delta \bm{q}_{220}^{\rm inh}
+
\delta \bm{q}_{220}^{(6)},
\qquad
\left\|\delta \bm{q}_{220}^{\rm inh}\right\|
\lesssim 10^{-14},
\end{equation*}
with no analogous cosmological suppression for $\delta \bm{q}_{220}^{(6)}$. This is the precise sense in which post-GW170817 cosmology collapses the homogeneous odd branch without collapsing the anisotropic completion of the same EFT\@.

\subsection{Literature-calibrated odd-parity response kernels}
\label{subsec:lit_calibrated_odd}

The stealth branch fixes the inherited odd-sector direction. For the anisotropy-activated direction we use the explicit QNM calculations of Ref.~\cite{MukohyamaTakahashiTomikawaYingcharoenrat2023}, avoiding ad hoc response matrices. The two Hayward parameters must be kept distinct. The asymptotically luminal bridge analysis of Sec.~\ref{subsec:luminality_loophole} uses
\begin{equation*}
\eta_{\rm BH}\equiv \sigma^3/r_g^3,
\end{equation*}
whereas the odd-parity QNM paper tabulates frequencies in terms of
\begin{equation*}
\eta_{\rm QNM}\equiv \sigma^3/\mu^3 = \hat\sigma^3.
\end{equation*}
Using
\begin{equation*}
\mu = r_g\,\frac{1+\eta_{\rm BH}}{1-4\eta_{\rm BH}-2\eta_{\rm BH}^2},
\end{equation*}
the two notations are related by
\begin{equation*}
\begin{aligned}
\eta_{\rm QNM}
&=
\eta_{\rm BH}
\left(
\frac{1-4\eta_{\rm BH}-2\eta_{\rm BH}^2}{1+\eta_{\rm BH}}
\right)^3
\\
&=
\eta_{\rm BH}-15\eta_{\rm BH}^2+\Order{\eta_{\rm BH}^3}.
\end{aligned}
\end{equation*}
Thus the two parameters agree only at leading order near the Schwarzschild point. On the positive branch, the stable asymptotically luminal interval of Eq.~\eqref{eq:eta_range} maps to
\begin{equation*}
0<\eta_{\rm QNM}\lesssim 2.17\times10^{-2},
\qquad
0<\hat\sigma\lesssim 0.279.
\end{equation*}
Among the positive direct-integration points quoted below, \(\hat\sigma=0.2\) lies inside this demonstrated interval, whereas \(\hat\sigma=0.4,0.6,0.8,1.0\) are retained only as literature-calibrated proxy continuation points used to show how the odd-sector response continues in the same direction beyond the explicitly stable asymptotically luminal window. We therefore use the Hayward branch in two distinct ways: Eq.~\eqref{eq:alphat_hayward} gives an explicit asymptotically luminal realization of Proposition~1, while the direct-integration frequencies calibrate the size and orientation of the open odd-sector branch.

With this notation understood, the small-\(\eta_{\rm QNM}\) expansion of Ref.~\cite{MukohyamaTakahashiTomikawaYingcharoenrat2023} on the Hayward timelike-scalar background gives
\begin{align*}
\mu \omega_{220}
={}&
(0.747343-0.177925\,\ii)
\\
&-
(0.507532+0.0372539\,\ii)\eta_{\rm QNM}
\\
&+
(1.17402+0.491105\,\ii)\eta_{\rm QNM}^2
+\Order{\eta_{\rm QNM}^3}.
\end{align*}
The leading anisotropy-activated odd-sector response in the dominant-mode plane is therefore
\begin{equation}
\delta \bm{q}_{220}^{(6)}
\equiv
\begin{pmatrix}
\delta q_{220}^{\rm R} \\
\delta q_{220}^{\rm I}
\end{pmatrix}
=
\begin{pmatrix}
-0.679\\
+0.209
\end{pmatrix}\eta_{\rm QNM}
+\Order{\eta_{\rm QNM}^2},
\label{eq:response_eta}
\end{equation}
Equation~\eqref{eq:response_eta} replaces a schematic response coefficient by a published odd-sector strong-field calibration.

\begin{figure}[t]
\centering
\subfloat[Parameter conversion and explicit admissible interval.\label{fig:hayward_interval}]{%
  \includegraphics[width=0.485\textwidth]{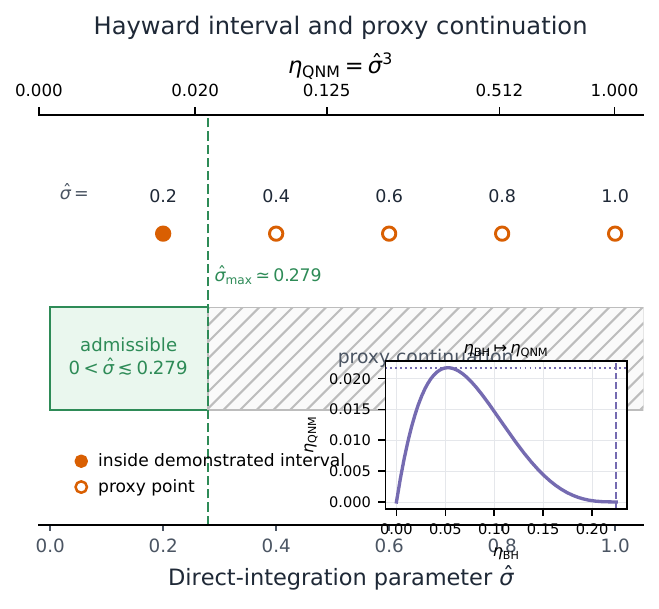}}
\hfill
\subfloat[Odd-sector footprints in the dominant-mode plane.\label{fig:hayward_plane}]{%
  \includegraphics[width=0.485\textwidth]{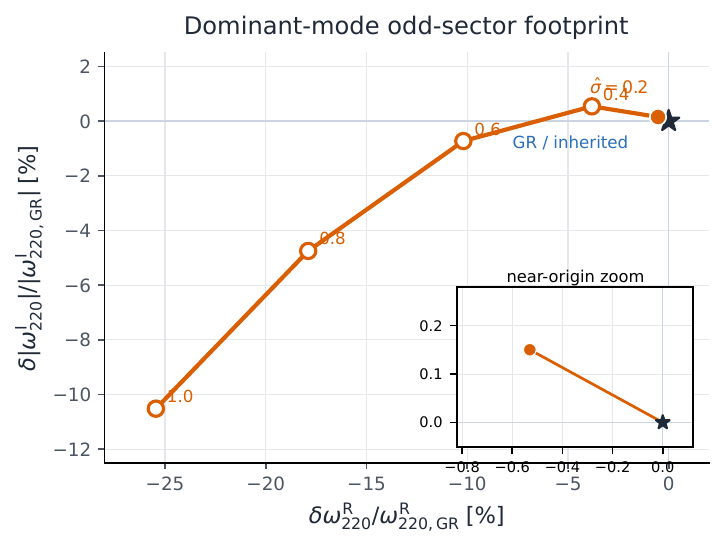}}
\caption{Hayward calibration of the anisotropy-activated odd sector. Panel (a) disentangles the black-hole background parameter \(\eta_{\rm BH}\) from the QNM parameter \(\eta_{\rm QNM}=\hat\sigma^3\) and marks the demonstrated positive stable asymptotically luminal interval. The filled marker at \(\hat\sigma=0.2\) lies inside that interval, while the larger positive points are shown explicitly as proxy continuation beyond it. Panel (b) translates the same information into the dominant-mode spectroscopy plane. The admissible point remains close to the GR origin, whereas the proxy continuation reveals the direction in which the anisotropy-activated odd branch pulls the real and imaginary parts of the \(220\) mode. Read together, the two panels separate what is demonstrably allowed from what is only literature-calibrated continuation, while still making the geometry of the open odd-sector branch immediately visible.}
\label{fig:hayward_panels}
\end{figure}

The same reference gives direct-integration values for the fundamental mode beyond the linear regime. With $\mu\omega_{220}^{\rm GR}=0.747343-0.177925\,\ii$, the literature-calibrated footprints are listed in Table~\ref{tab:hayward220} and plotted in Fig.~\ref{fig:hayward_panels}(b). Their geometry is useful: the real part shifts substantially negative, while the damping-rate shift is milder and can change sign along the branch.

\begin{table}[H]
\caption{Literature-calibrated odd-parity footprints from direct integration on the Hayward timelike-scalar background of Ref.~\cite{MukohyamaTakahashiTomikawaYingcharoenrat2023}. We quote the complex fundamental frequency and the induced shifts relative to $\mu\omega_{220}^{\rm GR}=0.747343-0.177925\,\ii$. The first positive row, \(\hat\sigma=0.2\), lies inside the explicit stable asymptotically luminal interval \(0<\hat\sigma\lesssim 0.279\) obtained by translating Eq.~\eqref{eq:eta_range}. The rows \(\hat\sigma=0.4,0.6,0.8,1.0\) are shown as literature-calibrated proxy continuation points outside that explicit interval.}
\label{tab:hayward220}
\begin{tabular}{c c c c}
\(\hat\sigma\) & \(\mu \omega_{220}\) & \(\delta \omega^{\rm R}_{220}/\omega^{\rm R}_{220,{\rm GR}}\) & \(\delta |\omega^{\rm I}_{220}|/|\omega^{\rm I}_{220,{\rm GR}}|\) \\
\hline
0.2 & \(0.743356 - 0.178193\,\ii\) & \(-0.53\%\) & \(+0.15\%\) \\
0.4 & \(0.718887 - 0.178884\,\ii\) & \(-3.81\%\) & \(+0.54\%\) \\
0.6 & \(0.671222 - 0.176619\,\ii\) & \(-10.19\%\) & \(-0.73\%\) \\
0.8 & \(0.613627 - 0.169466\,\ii\) & \(-17.89\%\) & \(-4.75\%\) \\
1.0 & \(0.557171 - 0.159204\,\ii\) & \(-25.45\%\) & \(-10.52\%\) \\
\end{tabular}
\end{table}

Figure~\ref{fig:hayward_panels} summarizes the strong-field lesson. Only \(\hat\sigma=0.2\) lies inside the demonstrated positive stable asymptotically luminal interval; the larger points trace proxy continuation along the same response direction. Even one admissible point is enough to fix the orientation of the open odd-sector branch in the \((\delta\omega^{\rm R}_{220},\delta|\omega^{\rm I}_{220}|)\) plane, separated from the inherited branch near the GR origin.

\subsection{Detector-facing hierarchy}
\label{subsec:detector_hierarchy_concrete}

The explicit odd-sector results become observationally meaningful once they are paired with the Bayesian detector covariances of Table~\ref{tab:ringdown_bayes}. The inherited branch remains the clean null benchmark. Equation~\eqref{eq:inherited_bound_final} shows that, once the remnant mass is fixed externally, the odd-parity stealth branch is suppressed below the $10^{-14}$ level; without such a calibration it is exactly degenerate with the GR mass parameter and therefore invisible in odd-parity ringdown. The open target is the anisotropy-activated direction calibrated in Sec.~\ref{subsec:lit_calibrated_odd}.

Writing Eq.~\eqref{eq:response_eta} as
\(\delta\bm q_{220}^{(6)}=\bm W_{\eta}\,\eta_{\rm QNM}+\Order{\eta_{\rm QNM}^2}\)
with
\(\bm W_{\eta}=(-0.679,\,0.209)^{\mathsf T}\),
we model the admissible and proxy-calibrated priors by
\begin{align}
\mathbf C_{\rm prior}^{\rm adm}
&=
\left(\frac{\eta_{\rm max}}{2}\right)^2
\bm W_{\eta}\bm W_{\eta}^{\mathsf T}
+\sigma_{\perp}^2\,\hat{\bm n}_{\perp}\hat{\bm n}_{\perp}^{\mathsf T},
\nonumber\\
\mathbf C_{\rm prior}^{\rm proxy}
&=
(0.03)^2
\bm W_{\eta}\bm W_{\eta}^{\mathsf T}
+\sigma_{\perp}^2\,\hat{\bm n}_{\perp}\hat{\bm n}_{\perp}^{\mathsf T},
\label{eq:prior_models_eta}
\end{align}
where
\(\eta_{\rm max}=2.17\times10^{-2}\)
encodes the demonstrated positive asymptotically luminal interval,
\(\hat{\bm n}_{\perp}\cdot\bm W_{\eta}=0\),
and
\(\sigma_{\perp}=2\times10^{-4}\)
keeps the theory support narrow in the orthogonal direction. The first covariance is a centered Gaussian surrogate for the demonstrated admissible interval; the second is broader and is used only to ask when the same odd-sector response direction would become observable if the branch extended beyond the currently demonstrated window.

Projecting the Bayesian detector covariance products of Table~\ref{tab:ringdown_bayes} and the machine-readable summary CSV onto the response direction \(\bm W_{\eta}\) gives effective one-parameter resolutions \(\sigma(\eta)=(2.47\%,\,1.53\%,\,0.55\%)\) for the current-like, ET-like, and LISA-like reference rows. For comparison, the admissible endpoint at \(\eta_{\rm max}\) corresponds to a 1.54\% displacement along the same direction, while the direct-integration point \(\hat\sigma=0.2\) gives only 0.55\%. The proxy continuation reaches 3.80\% already at \(\hat\sigma=0.4\), then continues to 9.52\%, 15.7\%, and 21.2\% for \(\hat\sigma=0.6,0.8,1.0\). In the detector-whitened language of Sec.~\ref{sec:phenom}, the largest generalized eigenvalue is
\begin{equation*}
\begin{aligned}
\lambda_1^{\rm adm}&=(0.19,\,0.51,\,3.93),\\
\lambda_1^{\rm proxy}&=(1.48,\,3.87,\,30.1),
\end{aligned}
\end{equation*}
for the current-like, ET-like, and LISA-like scenarios in that order.

In the compact Bayesian injection used here, the demonstrated admissible interval remains below threshold for the current-like and ET-like references but reaches the aggressive LISA-like covariance. The broader proxy continuation crosses threshold already for the current-like reference and is accessible for ET-like and LISA-like covariances. This is a detector-layer sensitivity statement, not a real-data detection claim: the injection and start-time tests show where the anisotropy-activated direction becomes measurable under controlled waveform assumptions.

\begin{figure}[t]
\centering
\subfloat[Bayesian 90\% detector ellipses in the dominant-mode plane, overlaid with the admissible prior and the broader proxy prior.\label{fig:detectability_regions}]{%
  \includegraphics[width=0.485\textwidth]{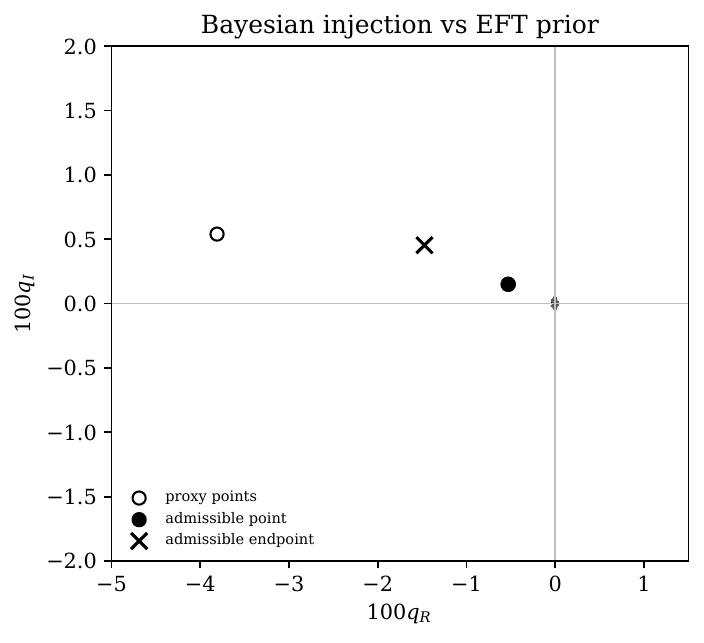}}
\hfill
\subfloat[Detector-weighted reach in the leading odd-sector consistency mode.\label{fig:odd_shift_hierarchy}]{%
  \includegraphics[width=0.485\textwidth]{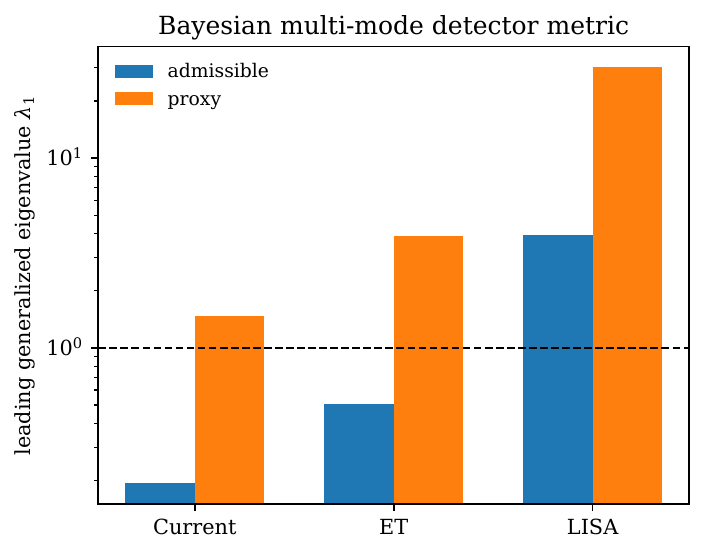}}
\caption{Detector-facing consequences of the worked odd-sector branch. Panel~(a) overlays the Bayesian 90\% detector ellipses derived from Eq.~\eqref{eq:bayes_injection_waveform} and the covariance products associated with Table~\ref{tab:ringdown_bayes} on the admissible and proxy-calibrated odd-sector priors of Eq.~\eqref{eq:prior_models_eta}. The filled point marks the admissible direct-integration value at \(\hat\sigma=0.2\); the open points trace the broader proxy continuation. Panel~(b) compresses the same comparison to the leading detector-whitened consistency eigenvalue. The admissible prior remains below threshold for the current-like and ET-like references but crosses in the aggressive LISA-like covariance, whereas the broader proxy prior crosses \(\lambda_1=1\) throughout this reference ladder.}
\label{fig:detectability_hierarchy}
\end{figure}

Figures~\ref{fig:hayward_panels} and~\ref{fig:detectability_hierarchy} give the quantitative forecast. The first fixes the anisotropy-activated response direction and separates the admissible point from proxy continuation. The second compares that direction with the Bayesian detector covariances: the explicit asymptotically luminal interval is hidden for the current-like and ET-like rows but visible in the aggressive LISA-like covariance, while the proxy continuation is a current-like/ET-like detector-reach target.

\begin{figure}[t]
\centering
\includegraphics[width=0.90\textwidth]{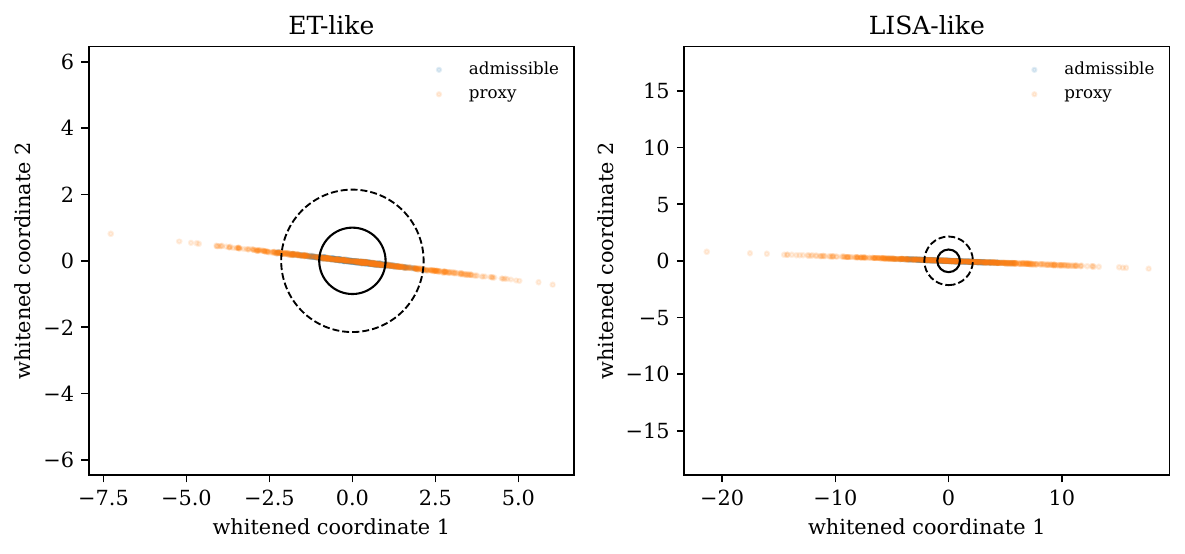}
\caption{Detector-whitened theory reach of the odd-sector prior. The same admissible and proxy covariances used in Fig.~\ref{fig:detectability_hierarchy} are shown in the leading consistency-coordinate plane after whitening by the ET-like and LISA-like Bayesian detector metrics. The solid circle marks the unit detectability threshold and the dashed circle gives a two-dimensional 90\% reference contour. Since the Hayward prior used here is a one-parameter pushforward, its support is rank one; the finite vertical thickness is drawn only to make the bands legible. The figure makes the threshold statement geometrically explicit: the admissible band stays inside the ET-like unit contour but crosses the aggressive LISA-like metric, while the proxy continuation crosses both metrics and extends well beyond the LISA-like threshold.}
\label{fig:theory_reach_modes}
\end{figure}

The whitened plot separates constraints from targets. The admissible band remains below threshold for current-like and ET-like covariances but enters the aggressive LISA-like target region. The proxy band is used only to motivate future sensitivity to the anisotropy-activated direction, not as a demonstrated exclusion region.

\subsection{Weak-mixing even-parity continuation}
\label{subsec:even_weakmix}

The timelike-scalar literature does not yet provide a generic even-parity QNM census comparable to the odd sector. It nevertheless contains enough structure to extract one robust result. Recent work on the even-parity EFT shows that, in a scordatura-compatible subclass around an approximately stealth Schwarzschild background, the scalar perturbation decouples from the metric perturbations at leading order in the scordatura effect for monopole perturbations \cite{MukohyamaTakahashiTomikawaYingcharoenrat2025}. That observation can be turned into a general statement about any even sector in which scalar-metric mixing is controlled by a small parameter \(\epsilon_{\rm mix}\).

\paragraph*{Proposition 2.}
Consider a gravitational-led even-parity mode whose frequency-domain equations can be written schematically as
\begin{equation*}
\begin{pmatrix}
\hat{\mathcal Z}(\omega) & \epsilon_{\rm mix}\hat{\mathcal C} \\
\epsilon_{\rm mix}\hat{\mathcal D} & \hat{\mathcal S}(\omega)
\end{pmatrix}
\binom{\Psi_g}{\Psi_s}=0,
\end{equation*}
with a nondegenerate gravitational root \(\hat{\mathcal Z}(\omega_Z^{(0)})\Psi_g^{(0)}=0\) and with \(\hat{\mathcal S}(\omega_Z^{(0)})\) invertible. Then the gravitational-led even-parity QNM receives no correction linear in \(\epsilon_{\rm mix}\); the leading shift is quadratic and is governed by the Schur-complement kernel
\begin{equation}
\bigl[\hat{\mathcal Z}(\omega)-\epsilon_{\rm mix}^2\hat\Sigma(\omega)\bigr]\Psi_g=0,
\qquad
\hat\Sigma\equiv \hat{\mathcal C}\,\hat{\mathcal S}^{-1}\hat{\mathcal D}.
\label{eq:even_schur}
\end{equation}
Using the standard left-right QNM bilinear form for non-self-adjoint perturbation theory then gives
\begin{equation}
\delta \omega_Z^{\rm even}
=
\epsilon_{\rm mix}^2
\frac{\langle\!\langle \widetilde\Psi_g^{(0)},\hat\Sigma(\omega_Z^{(0)})\Psi_g^{(0)}\rangle\!\rangle}
{\langle\!\langle \widetilde\Psi_g^{(0)},\partial_{\omega}\hat{\mathcal Z}(\omega_Z^{(0)})\Psi_g^{(0)}\rangle\!\rangle}
+\Order{\epsilon_{\rm mix}^3}.
\label{eq:even_weakmix_shift}
\end{equation}
A derivation is given in App.~\ref{app:extensions}.

With weak mixing and the scordatura regime enforced, the gravitational-led even branch is protected against \emph{linear} scalar contamination. The remaining question is the size of the quadratic response kernel \(\hat\Sigma\) for generic multipoles and viable branches. Proposition~2 does not replace a full even-sector posterior pushforward, but it puts the even sector on a controlled perturbative footing.

\subsection{Slow-spin continuation of the inherited branch}
\label{subsec:slowspin_inherited}

Recent EFT work on Kerr ringdown motivates asking which part of the present map already survives once the remnant spins \cite{MaenautEtAl2024,BoyceSantos2026}. A full timelike-scalar axisymmetric transport is still pending, but the inherited branch admits one immediate extension. Let \(\chi\equiv a/M\) be the dimensionless spin and write the GR Kerr spectrum as
\begin{equation*}
M\omega_{\ell mn}^{\rm Kerr}=F_{\ell mn}(\chi).
\end{equation*}
For the inherited isotropic branch studied above, the odd sector is controlled only by the constant tensor-speed deformation \(c_T^2=1+\alpha_T\). If the slow-spin continuation of this branch is obtained by the same uniform light-cone rescaling that produced Eq.~\eqref{eq:qnm_scaling_stealth}, then the dimensionless Kerr spectrum factorizes as
\begin{equation}
M\omega_{\ell mn}^{\rm inh}(\chi,\alpha_T)
=
(1+\alpha_T)^{3/2}F_{\ell mn}(\chi).
\label{eq:slowspin_factorization}
\end{equation}
Because \(\chi=a/M\) is dimensionless, it is unchanged by the common rescaling of the temporal and radial units on the inherited branch. Equation~\eqref{eq:slowspin_factorization} therefore implies
\begin{equation}
\frac{\delta\omega_{\ell mn}^{\rm inh}}{\omega_{\ell mn}^{\rm Kerr}(\chi)}
=
(1+\alpha_T)^{3/2}-1
=
\frac{3}{2}\alpha_T+\Order{\alpha_T^2},
\label{eq:slowspin_relative_shift}
\end{equation}
independently of \(\chi\) within the domain of validity of the inherited-isotropic continuation. As in the Schwarzschild case, this common factor can be absorbed into a mass redefinition \(M\to M_{\rm eff}=M(1+\alpha_T)^{-3/2}\) at fixed \(\chi\). The slow-spin inherited branch is therefore again odd-sector mass-degenerate; Eq.~\eqref{eq:slowspin_relative_shift} gives the residual rescaling after the mass scale is fixed externally. Inserting Eq.~\eqref{eq:alphaT_bound_explicit} gives the corresponding slow-spin inherited bound
\begin{equation}
\left|\frac{\delta\omega_{\ell mn}^{\rm inh}}{\omega_{\ell mn}^{\rm Kerr}(\chi)}\right|
\lesssim 9\times10^{-15}.
\label{eq:slowspin_inherited_bound}
\end{equation}
This is a modest extension, not a generic Kerr EFT spectrum. It shows that, if the inherited branch continues to slow rotation through the same isotropic light-cone rescaling, the inherited null result is not confined to exactly Schwarzschild remnants. The open Kerr problem is the spin-activated anisotropy sector.

\section{Discussion and outlook}
\label{sec:discussion}

The results are conditional on a chosen parent class and lift prescription. With those choices fixed, cosmological viability does not predict a unique ringdown spectrum; it instead carves the black-hole EFT down to a structured subset of spectroscopy space and isolates the anisotropy-activated directions that can survive post-GW170817 bounds.

\subsection{Interpretation of the consistency map}
\label{subsec:interpretation}

A compact way to state the result is probabilistic. Let
\begin{equation}
\mathfrak{C}
\equiv
\left\{
\begin{aligned}
&{\rm parent\ class},\,N,\,r_m,\,{\rm frame},\\
&{\rm background\ branch},\,{\rm observable\ basis}
\end{aligned}
\right\}
\label{eq:Cclass}
\end{equation}
collect the structural choices. The spectroscopy prior generated by Secs.~\ref{sec:map} and~\ref{sec:phenom} is then
\begin{equation}
\begin{aligned}
P(\delta \bm{q}_{\rm RD}\,|\,{\rm cosmo},\mathfrak{C})
={}&
\int \dd \bm{y}_*\,
P(\delta \bm{q}_{\rm RD}\,|\,\bm{y}_*,\mathfrak{C})\\
&\times
P(\bm{y}_*\,|\,{\rm cosmo},\mathfrak{C})
\end{aligned}
\label{eq:conditional_spectro_prior}
\end{equation}
where \(\bm{y}_*\) contains the cosmological data vector at the merger epoch together with the nuisance quantities introduced by the lift and transport steps. Thus the map is not a model-independent parametrization of every possible beyond-GR ringdown signal. It is a consistency relation between two background projections of the same EFT, conditional on a specified parent class and implementation.

Within that conditional framework, a future ringdown measurement \(\widehat{\delta\bm{q}}_{\rm RD}\) has three natural readings. If
\begin{equation*}
\widehat{\delta\bm{q}}_{\rm RD}
\in
\mathcal{D}^{(X)}_{p}\cap \mathcal{A}_{p},
\end{equation*}
the signal is both cosmology compatible and detectable for detector concept \(X\), and one has evidence for a viable non-GR strong-field imprint inside the class \(\mathfrak{C}\). If
\begin{equation}
\widehat{\delta\bm{q}}_{\rm RD}
\in
\mathcal{N}_{p},
\label{eq:excluded_case}
\end{equation}
the data exclude the chosen class \(\mathfrak{C}\) at the stated confidence level, though not modified gravity in general. If
\begin{equation*}
\widehat{\delta\bm{q}}_{\rm RD}
\in
\mathcal{U}^{(X)}_{p},
\end{equation*}
the theory remains viable but the detector is not informative enough along that direction. The no-go statements are correspondingly restricted: they rule out portions of ringdown parameter space for a fixed and explicitly stated EFT ancestry, not arbitrary UV completions or unrelated strong-field deformations.

This also sharpens the positive result. Cosmological luminality does \emph{not} force black-hole ringdown to be GR-like. It suppresses the isotropic FLRW branch while leaving room for anisotropy-activated near-zone operators, with the \(M_6\) sector in Secs.~\ref{subsec:arbbackground},~\ref{subsec:allowed_regions}, and~\ref{subsec:lit_calibrated_odd} as the prototype. The observational target is therefore not an arbitrary beyond-GR deviation, but the smaller set of strong-field signatures that survive cosmological, background, and perturbative consistency filters at once.

\subsection{Robustness diagnostics and stress tests}
\label{subsec:robustness}

The provenance tables separate computed results from calibrated proxies and deferred extensions. This subsection assesses the stability of the computed pieces.

Because the result is conditional on the choices in Eq.~\eqref{eq:Cclass}, nominal allowed regions should be accompanied by stress tests. Let \(\xi\in\Xi_{\rm stress}\) denote a discrete or continuous implementation choice --- for example the jet order \(N\), the overlap radius \(r_m\), the frame used to evaluate the truncated expression, the cosmological-posterior parametrization, or the observable projection defining the spectroscopy plane. The corresponding prediction is
\begin{equation*}
\delta \bm{q}_{\rm RD}^{(\xi)}
=
\mathcal{Q}\circ \mathcal{M}_{\xi}[\bm{y}_*],
\end{equation*}
where \(\mathcal{M}_{\xi}\) abbreviates the chain from the cosmological EFT to black-hole observables.

A useful detector-weighted measure of robustness is the maximum drift between two such implementations,
\begin{equation}
\begin{aligned}
\Delta_{X}^{2}(\xi,\xi')
\equiv{}&
\sup_{\bm{y}_*\in\mathcal{A}_{\rm cos}}
\Bigl[
\delta \bm{q}_{\rm RD}^{(\xi)}-\delta \bm{q}_{\rm RD}^{(\xi')}
\Bigr]^{\mathsf T}\\
&\times
\bigl(\mathbf{C}_{\rm det}^{(X)}\bigr)^{-1}
\Bigl[
\delta \bm{q}_{\rm RD}^{(\xi)}-\delta \bm{q}_{\rm RD}^{(\xi')}
\Bigr]
\end{aligned}
\label{eq:Delta_robust}
\end{equation}
By construction, \(\Delta_X\ll 1\) means that the implementation ambiguity is smaller than the detector resolution, while \(\Delta_X\gtrsim 1\) means that the nominal forecast is not stable at the level relevant for actual measurements. Equation~\eqref{eq:Delta_robust} packages several otherwise separate theory systematics into a common observational unit.

The minimum set of diagnostics that should accompany any full implementation is then
\begin{align}
\Delta_{N}^{(X)}
&\equiv
\Delta_{X}(N+1,N),
\label{eq:DeltaN}
\\
\Delta_{r_m}^{(X)}
&\equiv
\Delta_{X}(r_m+\delta r_m,r_m),\notag
\\
\Delta_{\rm frame}^{(X)}
&\equiv
\Delta_{X}(E,J),\notag
\\
\Delta_{\rm prior}^{(X)}
&\equiv
\Delta_{X}({\rm samples},{\rm Gaussian}),\notag
\\
\Delta_{\rm obs}^{(X)}
&\equiv
\Delta_{X}(\Pi_{220+330},\Pi_{220}),
\label{eq:Deltaobs}
\end{align}
where the labels indicate, respectively, the lift-truncation drift, the overlap-matching drift, the frame-covariance drift, the posterior-modeling drift, and the observable-basis drift. The third of these is especially important: in the exact theory the physical spectrum should be independent of whether the computation is organized in the Jordan frame or in the almost Einstein frame, so a nonzero \(\Delta_{\rm frame}^{(X)}\) is a direct measure of truncation or projection artifacts rather than of genuine physics.

Equations~\eqref{eq:DeltaN}--\eqref{eq:Deltaobs} allow one to define a robust allowed region by intersection over the stress-test family,
\begin{align}
\mathcal{A}_{p}^{\rm rob}
&\equiv
\bigcap_{\xi\in\Xi_{\rm stress}} \mathcal{A}_{p}(\xi),
\label{eq:Arob}
\\
\mathcal{D}_{p}^{(X),\rm rob}
&\equiv
\bigcap_{\xi\in\Xi_{\rm stress}} \mathcal{D}_{p}^{(X)}(\xi),\notag
\\
\mathcal{U}_{p}^{(X),\rm rob}
&\equiv
\mathcal{A}_{p}^{\rm rob}\setminus \mathcal{D}_{p}^{(X),\rm rob},
\qquad
\mathcal{N}_{p}^{\rm rob}
\equiv
\mathbb{R}^{n_q}\setminus \mathcal{A}_{p}^{\rm rob}.
\label{eq:UrobNrob}
\end{align}
These sets, not nominal contours alone, are the appropriate objects for falsifiable statements. A useful summary statistic is
\begin{equation*}
\Delta_{\rm max}^{(X)}
\equiv
\max\Bigl\{
\Delta_N^{(X)},\,
\Delta_{r_m}^{(X)},\,
\Delta_{\rm frame}^{(X)},\,
\Delta_{\rm prior}^{(X)},\,
\Delta_{\rm obs}^{(X)}
\Bigr\},
\end{equation*}
and the forecast is observationally stable for detector \(X\) only when \(\Delta_{\rm max}^{(X)}\lesssim 1\). The criterion is conservative: the full theory-side ambiguity must remain subdominant to the detector uncertainty in the spectroscopy basis under consideration.

Table~\ref{tab:robustness_scorecard} records the robustness checks available in the present implementation. In the exact inherited branch, the lift-order, matching-radius, and frame drifts vanish in the odd sector because the spectrum is controlled by a single constant rescaling. For the anisotropy-activated Hayward direction, the direct QNM calculation is imported from the literature in one frame, so a true Jordan-versus-Einstein QNM comparison remains deferred. The table separates those cases rather than folding them into one nominal contour.

\begin{table}[H]
\caption{Robustness scorecard for the present benchmark implementation. Values are quoted in the detector-whitened ET-like metric when a numerical value is meaningful. ``Executed'' means that the diagnostic enters the claims made in this paper; ``deferred'' means that a dedicated calculation is required before the corresponding systematic can be included in a final observational contour.}
\label{tab:robustness_scorecard}
\small
\begin{tabular*}{\textwidth}{@{\extracolsep{\fill}}llll@{}}
\hline\hline
\parbox[t]{0.13\textwidth}{Diagnostic} & \parbox[t]{0.17\textwidth}{Sector tested here} & \parbox[t]{0.16\textwidth}{ET-like value/status} & \parbox[t]{0.45\textwidth}{Interpretation} \\
\hline
\parbox[t]{0.13\textwidth}{\(\Delta_N^{(X)}\)} & \parbox[t]{0.17\textwidth}{inherited odd branch} & \parbox[t]{0.16\textwidth}{0} & \parbox[t]{0.45\textwidth}{Exact scaling depends only on \(\alpha_T\), so increasing the jet order does not change \(\delta\omega_{220}\).} \\
\parbox[t]{0.13\textwidth}{\(\Delta_N^{(X)}\)} & \parbox[t]{0.17\textwidth}{\(N=1\) lift surrogate} & \parbox[t]{0.16\textwidth}{0.36} & \parbox[t]{0.45\textwidth}{Drift between \(\lambda_{\rm reg}=0.1\) and 10 after projection along \(\bm W_\eta\); sub-threshold.} \\
\parbox[t]{0.13\textwidth}{\(\Delta_{r_m}^{(X)}\)} & \parbox[t]{0.17\textwidth}{inherited odd branch} & \parbox[t]{0.16\textwidth}{0} & \parbox[t]{0.45\textwidth}{No nontrivial radial profile enters the constant inherited rescaling.} \\
\parbox[t]{0.13\textwidth}{\(\Delta_{\rm prior}^{(X)}\)} & \parbox[t]{0.17\textwidth}{admissible vs.\ proxy priors} & \parbox[t]{0.16\textwidth}{reported separately} & \parbox[t]{0.45\textwidth}{The demonstrated admissible interval crosses only in the aggressive LISA-like metric; the proxy continuation is reported separately and must not be read as an established viability region.} \\
\parbox[t]{0.13\textwidth}{\(\Delta_{\rm frame}^{(X)}\)} & \parbox[t]{0.17\textwidth}{inherited odd branch} & \parbox[t]{0.16\textwidth}{0} & \parbox[t]{0.45\textwidth}{Dimensionless odd-mode ratios are invariant under the constant inherited rescaling.} \\
\parbox[t]{0.13\textwidth}{\(\Delta_{\rm frame}^{(X)}\)} & \parbox[t]{0.17\textwidth}{Hayward proxy} & \parbox[t]{0.16\textwidth}{deferred} & \parbox[t]{0.45\textwidth}{Requires a direct Jordan/E-frame QNM comparison, not performed here.} \\
\parbox[t]{0.13\textwidth}{\(\Delta_{\rm obs}^{(X)}\)} & \parbox[t]{0.17\textwidth}{Bayesian multi-mode injection} & \parbox[t]{0.16\textwidth}{0.25} & \parbox[t]{0.45\textwidth}{Start-time and mode-content variation are quantified by Figs.~\ref{fig:bayes_ringdown_posteriors}--\ref{fig:start_time_systematics}; population and real-noise systematics remain deferred.} \\
\hline\hline
\end{tabular*}
\end{table}

The resulting quantitative statement is narrow but robust. In the tested odd sector the inherited branch is exactly frozen and mass-degenerate. The demonstrated admissible Hayward interval remains below the current-like and ET-like thresholds but reaches the aggressive LISA-like target region in the compact Bayesian detector ladder. The broader proxy continuation is useful for detector reach; it is not counted as a demonstrated cosmology-conditioned exclusion region.

\subsection{Present limitations of the construction}
\label{subsec:limitations}

\paragraph*{Direct-footprint scope versus controlled extensions.}
The only direct integration footprints reported in Sec.~\ref{sec:benchmarks} are still odd-parity footprints, and that remains the numerically strongest part of the paper. The analysis nonetheless goes beyond a purely odd/nonspinning scope in two ways: Proposition~2 gives a weak-mixing even-sector protection result, and Eqs.~\eqref{eq:slowspin_factorization}--\eqref{eq:slowspin_inherited_bound} extend the inherited null branch to slow spin. Broader claims about full black-hole spectroscopy remain prospective.

\paragraph*{Proxy status of the Hayward calibration.}
The Hayward branch used in Sec.~\ref{subsec:lit_calibrated_odd} is a conservative strong-field proxy: published odd-parity QNM frequencies exist for it, and the asymptotically luminal branch can be written explicitly in EFT language. In the present analysis only the \(\hat\sigma=0.2\) point lies inside the demonstrated positive stable asymptotically luminal interval; the larger positive points are kept only to trace the same response direction beyond that window. Their role is to set the scale and orientation of the anisotropy-activated odd-sector response, not to stand in for a fully sampled cosmology-conditioned posterior.

\paragraph*{Finite-jet ambiguity and parent-class dependence.}
The controlled lift introduced in Sec.~\ref{subsec:lift} is a disciplined right-inverse of the FLRW projection, but it is not unique. Different regularizers, truncation orders, and parent classes can agree on cosmology while differing in the anisotropic completion activated near the black hole. FLRW data simply do not fully determine inhomogeneous strong-field operators. Any exclusion claim must therefore be stated class by class rather than advertised as a blanket exclusion of modified gravity.

\paragraph*{Restriction to static backgrounds, modulo the inherited slow-spin continuation.}
The transported EFT is built on static, spherically symmetric backgrounds and uses the overlap region of Eq.~\eqref{eq:matching_shell}. Real merger remnants are rotating, and for high signal-to-noise events the spin dependence of the mode spectrum will be at least as important as the radial transport emphasized here. The inherited isotropic branch is not confined to the exactly Schwarzschild limit: Eqs.~\eqref{eq:slowspin_factorization}--\eqref{eq:slowspin_inherited_bound} show that its cosmological suppression survives under the stated inherited-isotropic slow-spin continuation. The genuinely open Kerr problem is the spin-activated anisotropy sector and the full axisymmetric transport, not the inherited null branch itself \cite{Dreyer2004,BertiCardosoWill2006,MaenautEtAl2024,BoyceSantos2026}.

\paragraph*{Odd/even parity asymmetry.}
The odd-parity sector is still much better controlled than the even-parity sector. The generalized Regge--Wheeler description is already mature, whereas the even sector still depends sensitively on scalar-metric mixing and, around stealth backgrounds, on the scordatura regulator introduced to avoid strong coupling \cite{MukohyamaTakahashiYingcharoenrat2022,MukohyamaTakahashiTomikawaYingcharoenrat2025}. Proposition~2 shows that this asymmetry is not absolute: once the weak-mixing regime is enforced, the gravitational-led even sector is protected against linear contamination and the first correction is quadratic in the mixing parameter. The cleanest result remains the odd-sector consistency map, but the even sector now has a controlled perturbative foothold.

\paragraph*{Forecast idealizations.}
The detector comparison in Secs.~\ref{sec:phenom} and~\ref{sec:benchmarks} uses Bayesian time-domain injections, one-, two-, and three-mode recovery models, analytic amplitude marginalization, and explicit start-time variation. It is still a compact simulated likelihood, not a complete event analysis. A full observational forecast would need event-specific detector noise, calibration uncertainty, sky position and inclination, numerical-relativity-calibrated post-merger waveforms, selection effects, and a population model. Figure~\ref{fig:detectability_hierarchy} is therefore a self-consistent detector-weighted benchmark with an internal Bayesian systematic envelope, not a final LVK/LISA data product \cite{MaggioreEtAl2020,AmaroSeoaneEtAl2017,IsiFarr2021,VolkelDhani2025}.

\paragraph*{Astrophysical and environmental systematics.}
Several sources of contamination can still imitate or obscure small ringdown shifts: residual matter in the near environment, imperfect mode identification, beyond-linear perturbative effects, and errors inherited from the inspiral-merger-remnant inference chain \cite{IsiFarr2021,VolkelDhani2025}. These effects do not undermine the EFT map, but they set the scale at which spectroscopy becomes a clean consistency test. The robust regions of Eqs.~\eqref{eq:Arob}--\eqref{eq:UrobNrob} should ultimately be intersected with an astrophysical systematic budget before any strong-field inconsistency claim is made.

\subsection{Outlook and next steps}
\label{subsec:outlook}

The most urgent next step is a \emph{full} Kerr extension of the anisotropy-activated sector. The inherited branch is partly under control: Eqs.~\eqref{eq:slowspin_factorization}--\eqref{eq:slowspin_inherited_bound} show that its cosmological freezing survives the slow-spin continuation used here. The missing ingredients are axisymmetric transport of FLRW-silent operators and Kerr response kernels for the dominant rotating modes. Recent EFT progress on Kerr ringdown makes that program much more concrete \cite{MaenautEtAl2024,BoyceSantos2026}. On the data-analysis side, the compact Bayesian injection should be extended to event-specific multi-mode likelihoods, realistic detector noise curves, and hierarchical remnant populations.

A second step is to move from single-event forecasts to population-level inference. If \(\bm{\eta}\) denotes the shared EFT hyperparameters controlling the surviving branches of the map, the natural hierarchical posterior is
\begin{equation}
P(\bm{\eta},\mathfrak{C}\,|\,\{d_i\})
\propto
P(\bm{\eta},\mathfrak{C})
\prod_i
\int \dd \vartheta_i\,
\mathcal{L}_i\bigl(d_i\,|\,\vartheta_i,\bm{\eta},\mathfrak{C}\bigr),
\label{eq:hierarchical_posterior}
\end{equation}
with event-specific nuisance parameters \(\vartheta_i\) for remnant mass, spin, inclination, and mode amplitudes. The strength of this approach is cumulative: it lies in combining a cosmology-conditioned prior with many black-hole mergers, not in extracting all information from one spectacular event.

The same logic should also generalize beyond the scalar-tensor classes used here. Vector-tensor dark-energy EFTs, parity-violating sectors, and multi-field completions are natural next targets, provided that a sufficiently explicit arbitrary-background perturbation EFT is available on the black-hole side. In each case the structural question is the same: which operator combinations are already visible on homogeneous cosmological backgrounds, which are silent on FLRW but activated near a black hole, and which remain genuinely unconstrained strong-field data?

A mature implementation should also release cosmology-conditioned response kernels, consistency-eigenmode spectra for benchmark detector concepts, and robust region charts such as Eqs.~\eqref{eq:Arob}--\eqref{eq:UrobNrob}. These data products would let ringdown analyses use cosmological viability information without rebuilding the EFT transport each time.

\section{Conclusions}
\label{sec:conclusions}

We have constructed an EFT consistency map that carries cosmologically viable scalar-tensor gravity to black-hole ringdown observables without starting from a theory-by-theory strong-field calculation. The chain
\begin{equation*}
\begin{aligned}
\Theta_{\rm DE}^{J}(t_*)
&\longrightarrow J^{N}\mathcal{F}
\longrightarrow \mathcal{C}_{A}^{E}(r)\\
&\longrightarrow \delta \bm{q}_{\rm RD}
\end{aligned}
\end{equation*}
turns the problem into a sequence of explicit steps: identify the cosmologically allowed branch, lift it to a controlled covariant jet, transport it to the black-hole EFT, and project it onto parity-resolved ringdown observables. Black-hole spectroscopy then becomes a consistency test linking two background projections of the same parent theory.

The main physics lesson is straightforward. Cosmological viability does not collapse the strong-field sector to GR; it constrains the homogeneous and isotropic projection of the EFT\@. When the same theory is evaluated on an anisotropic black-hole background, additional operator combinations can become active. For the luminal shift-symmetric stealth-Schwarzschild branch this distinction yields an exact null result: the inherited odd sector obeys $r_H\omega_{\ell n}=r_H\omega_{\ell n}^{\rm GR}(1+\alpha_T)^{3/2}$, is degenerate with a remnant-mass redefinition in odd parity, and is pushed below a fractional externally calibrated $(220)$ shift of $10^{-14}$ by the GW170817/GRB170817A luminality bound. The same null result survives the inherited-isotropic slow-spin continuation. Proposition~2 adds that, in the weak-mixing/scordatura regime, gravitational-led even modes receive no linear scalar-metric contamination.

The remaining window is the anisotropy-activated sector. The \(M_6\) branch realizes it explicitly: the operator is silent on FLRW, asymptotically luminal, and capable of producing finite odd-sector nonluminality in the near zone. That direction is calibrated using one positive point inside the demonstrated asymptotically luminal interval; the larger literature-calibrated points are kept as a proxy continuation; and the response is pushed through the compressed cosmological posterior generated by the ancillary workflow. Direct QNM integration for every posterior draw remains a next step. The resulting picture is a three-way split: an inherited branch that is mass-degenerate and cosmologically frozen, a weak-mixing even branch that is parametrically protected, and an anisotropy-activated branch that remains the natural spectroscopy target. Together, the finite-jet lift, arbitrary-background EFT embedding, frame bridge, Bayesian multi-mode injections, and start-time tests provide a concrete path from cosmological modified gravity to black-hole spectroscopy.

\appendix

\section{Compact technical derivations}
\label{app:lift}

This appendix collects the formulae needed to reproduce the covariance and response conventions used in the main text. The implementation checks are contained in the executable scripts distributed with the source package.

\subsection{Finite-jet lift covariance}
\label{app:lift_demo}

Linearizing the FLRW projection of a finite jet gives
\begin{equation}
\delta\bm y=\mathbf A\,\delta\bm f+\bm n,
\qquad
\bm n\sim {\cal N}(0,\mathbf C_y),
\label{eq:linearized_lift_map_appendix}
\end{equation}
where \(\bm f\) denotes the retained covariant jet coefficients. With a Gaussian structural prior of precision \(\lambda_{\rm reg}\mathbf R\), the posterior covariance is
\begin{equation}
\mathbf C_f=
\left(\mathbf A^{\mathsf T}\mathbf C_y^{-1}\mathbf A+
\lambda_{\rm reg}\mathbf R\right)^{-1}.
\label{eq:Cf_appendix}
\end{equation}
This paper uses the Bayesian inverse-Hessian convention throughout. The deterministic Tikhonov estimator would instead propagate the data covariance as \(\mathbf L\mathbf C_y\mathbf L^{\mathsf T}\), with \(\mathbf L=(\mathbf A^{\mathsf T}\mathbf C_y^{-1}\mathbf A+\lambda_{\rm reg}\mathbf R)^{-1}\mathbf A^{\mathsf T}\mathbf C_y^{-1}\); that sandwich covariance is not the one plotted or pushed forward here. A singular-value decomposition
\begin{equation}
\mathbf A=\mathbf U\,{\rm diag}(s_i)\,\mathbf V^{\mathsf T}
\label{eq:svd_appendix}
\end{equation}
separates FLRW-image directions from null directions. The image modes are informed by cosmology, whereas the null modes are controlled only by the structural prior and by black-hole regularity conditions. The covariance of any black-hole EFT coefficient vector \(\bm c=\bm c_0+\mathbf B\delta\bm f\) is therefore
\begin{equation}
\mathbf C_c=\mathbf B\mathbf C_f\mathbf B^{\mathsf T}.
\label{eq:Cc_appendix}
\end{equation}

\subsection{Response kernels, robust regions, and Bayesian detector products}
\label{app:response}

For small deformations the ringdown response is written as
\begin{equation}
\delta\bm q_{\rm RD}=\mathbf R\,\delta\bm f,
\qquad
\mathbf C_{\rm RD}^{\rm cos}=\mathbf R\mathbf C_f\mathbf R^{\mathsf T}.
\label{eq:Rmatrix_appendix}
\end{equation}
Equivalently, in a continuous radial representation one may write
\begin{equation}
\delta q_i=\int_{r_H}^{\infty}K_{ia}(r)\,\delta c_a(r)\,\dd r,
\label{eq:kernel_continuous_appendix}
\end{equation}
which reduces to Eq.~\eqref{eq:Rmatrix_appendix} after discretizing the transported coefficient functions. The detector-weighted singular values of
\begin{equation}
\mathbf W=\mathbf C_{\rm det}^{-1/2}\mathbf R\mathbf C_f^{1/2}
\label{eq:svdW_appendix}
\end{equation}
are the consistency eigenmodes used in the detectability hierarchy. The robust allowed region quoted in Sec.~\ref{sec:discussion} is the intersection of the allowed regions obtained when the jet order, matching radius, frame convention, posterior compression, observable basis, recovery mode content, and ringdown start time are varied within the declared stress-test family.

The detector covariance used in Secs.~\ref{subsec:bayesian_injections} and~\ref{subsec:lit_calibrated_odd} is produced by the script \ringdowncode. The script injects the three-mode signal of Eq.~\eqref{eq:bayes_injection_waveform}, evaluates the whitened Gaussian likelihood on a two-dimensional \((q_R,q_I)\) grid, and analytically marginalizes over all linear sine/cosine amplitudes. Posterior means, covariance matrices, 90\% highest-posterior-density areas, maximum-a-posteriori grid points, and the generalized eigenvalues \(\lambda_1^{\rm adm}\) and \(\lambda_1^{\rm proxy}\) are written to \texttt{data/ringdown\_bayesian\_summary.csv}. The table in the main text is a direct \LaTeX{} export from the same file, and the posterior contours in Figs.~\ref{fig:bayes_ringdown_posteriors}--\ref{fig:start_time_systematics} are regenerated from the stored grids.

For a fixed grid point \((q_R,q_I)\) and a fixed start time, the waveform is linear in the sine/cosine amplitudes of the included modes. Writing the whitened data vector as \(\bm d\), the design matrix as \(\mathbf X(q_R,q_I)\), the white-noise variance as \(\sigma_n^2\), and the broad amplitude-prior scale as \(A_0\), the analytic marginal log-likelihood used in the ringdown script is
\begin{align}
\ln {\cal L}_{\rm marg}(q_R,q_I)
=&-\frac12\left[
\frac{\bm d^{\mathsf T}\bm d}{\sigma_n^2}
-\bm b^{\mathsf T}\mathbf S^{-1}\bm b
+\ln\det \mathbf S
\right]+{\rm const},
\nonumber\\
\mathbf S=&\frac{\mathbf X^{\mathsf T}\mathbf X}{\sigma_n^2}+A_0^{-2}\mathbf I,
\qquad
\bm b=\frac{\mathbf X^{\mathsf T}\bm d}{\sigma_n^2}.
\label{eq:app_ringdown_marginal_likelihood}
\end{align}

For reference, the Kerr baseline used to set the representative mode frequencies follows the standard fits of Ref.~\cite{BertiCardosoWill2006},
\begin{align}
M\omega^{\rm R}_{220,{\rm GR}}
&=1.5251-1.1568(1-\chi)^{0.1292},
\nonumber\\
Q_{220}
&=0.7000+1.4187(1-\chi)^{-0.4990},
\label{eq:fisher_fit_appendix}
\end{align}
with \(|\omega^{\rm I}_{220,{\rm GR}}|=\omega^{\rm R}_{220,{\rm GR}}/(2Q_{220})\). In the Bayesian injection this fit fixes only the representative scale and orientation of the dominant-mode response; the reported detector products are read from the marginalized likelihood rather than from a standalone single-mode Fisher calculation.

\subsection{Weak-mixing even sector and slow-spin continuation}
\label{app:extensions}

The even-sector protection statement follows from the block operator
\begin{equation}
\begin{pmatrix}
\hat{\mathcal Z}(\omega) & \epsilon_{\rm mix}\hat{\mathcal C} \\
\epsilon_{\rm mix}\hat{\mathcal D} & \hat{\mathcal S}(\omega)
\end{pmatrix}
\binom{\Psi_g}{\Psi_s}=0.
\label{eq:app_even_block_compact}
\end{equation}
If \(\hat{\mathcal S}\) is invertible at the gravitational-led root, eliminating \(\Psi_s\) gives
\begin{equation}
\left[\hat{\mathcal Z}(\omega)-
\epsilon_{\rm mix}^{2}\hat{\mathcal C}\hat{\mathcal S}^{-1}(\omega)\hat{\mathcal D}\right]
\Psi_g=0.
\label{eq:app_even_schur_compact}
\end{equation}
There is no term linear in \(\epsilon_{\rm mix}\): a gravitational-led mode must convert to the scalar sector and back before its root shifts. The resulting leading correction is Eq.~\eqref{eq:even_weakmix_shift} of the main text.

For the inherited slow-spin continuation, the only assumption is that \(\alpha_T\) enters the rotating inherited branch through the same homogeneous light-cone rescaling as in the Schwarzschild calculation. Then
\begin{equation}
M\omega_{\ell mn}^{\rm inh}(\chi,\alpha_T)
=(1+\alpha_T)^{3/2}
M\omega_{\ell mn}^{\rm Kerr}(\chi),
\label{eq:app_spin_factorization}
\end{equation}
so the low-redshift luminality bound gives \(|\delta\omega/\omega|\lesssim9\times10^{-15}\). Equation~\eqref{eq:app_spin_factorization} does not constrain spin-activated anisotropy sectors; those require the full axisymmetric transport problem.

\acknowledgments
AI-assisted tools were used for copy-editing and for drafting ancillary scripts. The authors checked and approved all scientific content, calculations, and conclusions.

\section*{Data and code availability}

The source archive contains the LaTeX source, the JCAP style file used for compilation, all figures, and the executable scripts \posteriorcode\ and \ringdowncode. The first script regenerates the compressed BAO-like, growth-like, and tensor-speed data, posterior samples, summary table, and posterior-pushforward figures stored under \texttt{data/} and \texttt{final\_figure\_pack/}. The second regenerates the Bayesian injection/recovery tables, posterior bank, and start-time-systematics figures. A companion script regenerates the finite-jet validation figures using the Bayesian inverse-Hessian covariance convention of Sec.~\ref{subsec:lift_demo} and App.~\ref{app:lift}. All numerical data are deterministic and seeded from the fiducial choices stated in Secs.~\ref{subsec:benchmark_inputs} and~\ref{subsec:bayesian_injections}. No proprietary survey data, external downloads, or non-public gravitational-wave data are required. On submission, the scripts and generated data products should be deposited as arXiv ancillary files or in a persistent public repository.

\end{document}